\documentclass{aa}
\usepackage{xcolor}
\usepackage{natbib}
\usepackage{graphicx}
\usepackage{stfloats}
\usepackage{txfonts}
\usepackage[breaklinks,colorlinks,urlcolor=blue,citecolor=blue,linkcolor=blue]{hyperref}
\begin{document}

   \title{Influence of planetary gas accretion on the shape and depth of gaps in protoplanetary discs}

   \author{C. Bergez-Casalou\inst{1}, B. Bitsch\inst{1}, A. Pierens\inst{2}, A. Crida\inst{3}, S. N. Raymond\inst{2}}

   \institute{\inst{1}Max-Planck-Institut für Astronomie, Königstuhl 17, 69117 Heidelberg, Germany, \email{bergez@mpia.de} \\ \inst{2}Laboratoire d’Astrophysique de Bordeaux, CNRS and Université de Bordeaux, Allée Geoffroy St. Hilaire, 33165 Pessac, France \\ \inst{3} Université Côte d’Azur / Observatoire de la Côte d’Azur — Lagrange (UMR 7293),
Boulevard de l’Observatoire, CS 34229, 06300 Nice, France }


 
 \abstract
  {It is widely known that giant planets have the capacity to open deep gaps in their natal gaseous protoplanetary discs. It is unclear, however, how gas accretion onto growing planets influences the shape and depth of their growing gaps. We performed isothermal hydrodynamical simulations with the Fargo-2D1D code, which assumes planets accreting gas within full discs that range from 0.1 to 260 AU. The gas accretion routine uses a sink cell approach, in which different accretion rates are used to cope with the broad range\ of gas accretion rates cited in the literature. We find that the planetary gas accretion rate increases for larger disc aspect ratios and greater viscosities. Our main results show that gas accretion has an important impact on the gap-opening mass: we find that when the disc responds slowly to a change in planetary mass (i.e., at low viscosity), the gap-opening mass scales with the planetary accretion rate, with a higher gas accretion rate resulting in a larger gap-opening mass. On the other hand, if the disc response time is short (i.e., at high viscosity), then gas accretion helps the planet carve a deep gap. As a consequence, higher planetary gas accretion rates result in smaller gap-opening masses. Our results have important implications for the derivation of planet masses from disc observations: depending on the planetary gas accretion rate, the derived masses from ALMA observations might be off by up to a factor of two. We discuss the consequences of the change in the gap-opening mass on the evolution of planetary systems based on the example of the grand tack scenario. Planetary gas accretion also impacts stellar gas accretion, where the influence is minimal due to the presence of a gas-accreting planet.
  
} 


   \authorrunning{C.Bergez-Casalou et al}

   \maketitle
%

\section{Introduction}

Recent ALMA observations have revealed protoplanetary discs with many diverse features in the gas \citep{Teague2018,Pinte2020} or in the dust \citep{Andrews2018}. An important question considers whether these features can be explained by the presence of planets. Recent exoplanet statistics show that most stars host planets around them \citep{Mayor2011,Howard2012,Suzuki2016,Mulders2018}, but it can be difficult to observe them directly as they might be extinct in the disc itself \citep{Sanchis2020}. A viable recourse is to rely on the impact of the presence of these planets on the disc. A number of studies have shown  that planets are capable of producing the features observed in these discs \citep{Pinilla2012}. For example, \cite{Zhang2018} shows that the gaps and rings seen in the DSHARP survey could be explained by the presence of gap-opening planets in the disc. 

The formation of gaps by giant planets in protoplanetary discs in the dust \citep{Paardekooper2006_dust} and in the gas \citep{LinPapaloizou1986,Crida2006,Fung2014,Kanagawa2015} has been studied in the past via hydrodynamical simulations. Different criteria for the gap-opening mass (i.e., planetary mass needed to open a gap of a given depth) have been derived as a function of disc parameters. The main motivation for these studies has, in fact, been the migration of the giant planets that have the capacity to open these deep gaps. However, it is only recently that these migration studies have started to factor in planetary gas accretion \citep{Durmann2015,Crida2017,Robert2018}.

Gas accretion is a complicated problem that requires 3D high resolution hydrodynamical simulations. The core accretion model \citep{Pollack1996} suggests that after forming a solid core, a gaseous envelope is slowly accreted until the planet reaches a critical mass where the mass of the gas envelope is equal to the mass of the core. The planet then enters a runaway growth phase of gas accretion during which we can assume that only gas is accreted \citep{Hubickyj2005,Lissauer2009}. Previous studies have described how gas accretion can be modeled in different frameworks: in 2D with and without migration  \citep{Crida2017,Kley1999} or in 3D \citep{Ayliffe2009,Machida2010,DAngelo2013,Lambrechts2019,Schulik2019}. These complex hydrodynamical simulations are needed to understand each phase of gas accretion. Due to their high computational cost, they are often integrated over short timescales. In fact, a lot of these recent simulations can only cover 100 planetary orbits (or less), making it impossible to study the long-term evolution of these systems within the 3D high resolution framework with the current computing systems.

As gas accretion requires a computationally expensive resolution, previous studies on gap-opening by giant planets neglected planetary gas accretion. Our goal in this paper is to apply established gas accretion rates \citep{Kley1999,Machida2010} in a 2D isothermal disc framework to study the long-term behavior of the growth of the planet and its influence on the shape and depth of the created gap.

This work is structured as follows. In Sect. \ref{section_numerics}, we describe our numerical set-up with the description of the FARGO-2D1D code \citep{Crida2007} and our gas accretion routine. In Sect. \ref{section_gasaccinflu}, we study the difference between an accreting planet and non-accreting planets, the impact that different accretion rate can have on the gap shape and on the accretion onto the star. We then explore different disc parameters ($h$ and $\alpha$) in Sect. \ref{section_influenceofparams}, as well as the impact of gas accretion on the gap-opening mass and migration of other planets that could form in these discs. We summarize our findings with a discussion of the consequences for observation and planetary system structures in Sect. \ref{section_discussion} and present our conclusions in Sect. \ref{section_conclusions}. 

\section{Numerical setup} \label{section_numerics}

\subsection{The FARGO-2D1D code} \label{section_setup}

\begin{figure}[!ht]
        \centering   
        \includegraphics[scale=1.0]{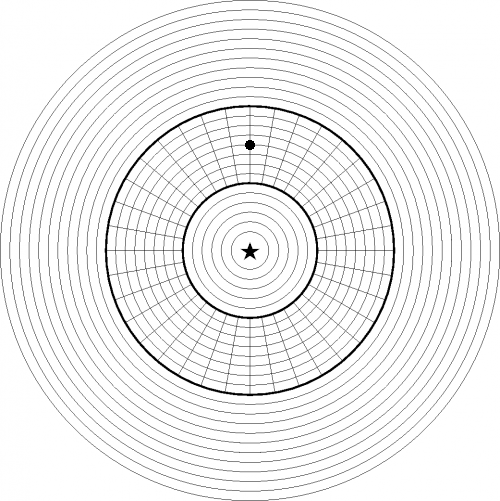}
        \caption{Sketch of the grid configuration in FARGO-2D1D. The 1D-grid ranges from 0.02 to 50.0 and the 2D grid from 0.15 to 4.5 in code units. The dot represents the location of the planet and the star represents the central star. The planet, located at $\rm r_{\rm p} = 1 = 5.2$ AU, is considered far enough from the boundaries of the 2D-grid for the disc to be considered axisymmetric.}
        \label{2D1D}
\end{figure}

In this paper, we simulate an accreting planet on a fixed circular orbit embedded in a viscous disc with the hydrodynamical code FARGO-2D1D \citep{Crida2007}. This code allows us to simulate the whole viscous evolution of a protoplanetary disc for a reasonable computational cost by surrounding a 2D grid ($\rm r,\phi$), where the planet is located, with two 1D grids that are azimuthally symmetric, as shown in figure \ref{2D1D}. The  extent of the grids is chosen so that the 1D grids are far enough from the planet that we can consider the disc to be axisymmetric. The 1D inner disc spans from $0.1$ AU to $0.78$ AU. The 2D grid then ranges from $0.78$ AU to $23.4$ AU, with a 1D outer disc from $23.4$ AU to $260$ AU. The resolution is such that there are five cells per Hill radius of the planet at t=0, which leads to $\rm N_r = 802$ and $\rm N_\phi=1158$ ($\rm dr/r_0 \simeq d\phi \simeq 0.005$). Considering that the resolution is fixed in time, the Hill sphere region will become more and more resolved as the planet grows ($\rm r_{\rm H} \propto m_{\rm p}^{1/3}$).

For computational accuracy, the code uses dimensionless units. We normalized masses with the mass of the central star, $\rm M_0 = M_\odot$ and lengths with the position of the planet, $\rm r_0 = 5.2$ AU and we set the gravitational constant as $\rm G=1$. We end up with a time normalized by the orbital time at $\rm r_0$, $\rm t_0 = (r_0^3/(GM_{*}))^{-1/2}$, meaning that the orbital period of the planet is $P = 2\pi t_0$.

The disc is locally isothermal and its aspect ratio $h=\rm H/r$ is constant. We investigate three different values for this ratio in Sect. \ref{section_diffh}, $ h=0.03$, $0.05$ and $0.07$. The density profile is defined by $\rm \Sigma (r) = \Sigma_0\times(r/r_0)^{-1}$ , where $\Sigma_0 = 3.10^{-4} = 93.6 \rm \; g/cm^2$ at $\rm r_0$. Its value is chosen such that the total mass of the disc is $\rm 0.1 M_{*}$, corresponding to a heavy disc with a large radius \citep{Baillie2019}. This large radial extent allows us to neglect self-gravity. The disc is subject to an $\alpha$-viscosity as described by \cite{Shakura1973}. We investigate the influence of the viscosity in Sect. \ref{section_diffvisco} by taking different values for $\alpha$ = $2\times10^{-2}$, $10^{-2}$, $10^{-3}$, and $10^{-4}$. Our fiducial setup is shown in the first row of Table \ref{table:1}, where we summarize the parameters of the different simulations we compare in this paper.

The accreting planet starts with a mass of $\rm m_{\rm p} = 20\;M_\oplus$ or $\rm m_{\rm p} = 10\;M_\oplus$, depending on the aspect ratio (see Table \ref{table:1}). With this initial mass, we can assume that the planet will be directly in the runaway gas accretion regime \citep{Pollack1996}. The accreting planet is introduced into the disc with a mass-taper function, making the planet grow from 0 to its initial mass in $n_{\rm orb} = 3$ orbits. The mass taper function is written as:
\begin{equation}
    m_{\rm taper} = \sin^2{(t/(4 n_{\rm orb}))}
    \label{masstapereq}
.\end{equation}
In order to let the disc adapt to the planet, we wait until the initial mass creates a steady gap. We define the equilibrium by a change of gap depth of less than $1\%$ within 100 orbits. The times needed to reach the steady state and steady gap depth are summarized in Table \ref{table:2} as a function of the disc parameters. In principle, the introduction of the planet in the simulation can change the resulting structure of the disc depending on the length of the ramp-up time \citep{Hammer2017}. However, this applies mostly for giant planets and low viscosities. For our initially small planets, we found no significant difference if we let gas accretion onto the planet start after a longer wait time.

\begin{table}            
\centering                          
\begin{tabular}{c c c c}        
\hline\hline                 
Section & $h$ & $\alpha$ & $\rm m_{init}$ \\    
\hline\hline                      
   3.X & 0.05 & $10^{-2}$ & $20\;M_\oplus$ \\     
   \hline
    & 0.03 & $10^{-2}$ & $20\;M_\oplus$ \\
   4.1 & 0.05 & $10^{-2}$ & $20\;M_\oplus$ \\
    & 0.07 & $10^{-2}$ & $20\;M_\oplus$ \\
   \hline
   4.2 & 0.05 & $2\times10^{-2}$; $10^{-2}$; $10^{-3}$; $10^{-4}$ & $20\;M_\oplus$ \\
   \hline
    & 0.05 & $2\times10^{-2}$; $10^{-2}$; $10^{-3}$; $10^{-4}$ & $20\;M_\oplus$ \\
    4.3; 4.4 & 0.07 & $10^{-2}$; $5\times10^{-3}$; $10^{-3}$; $10^{-4}$ & $20\;M_\oplus$ \\
    & 0.03 & $10^{-2}$; $10^{-3}$; $5\times10^{-4}$ & $10\;M_\oplus$ \\
\hline                                   
\end{tabular}
\caption{Explored disc parameters by sections. In Sect. \ref{section_gasaccinflu}, we use our fiducial parameters to study the influence of gas accretion. The influence of the aspect ratio is presented in Sect. \ref{section_diffh} and the different viscosities are studied in Sect. \ref{section_diffvisco}. In Sect. \ref{section_gapopeningmass}, we explored the gap-opening mass for the different aspect ratios and viscosities. As a $20\;M_\oplus$ already creates a deep gap in the low aspect ratio and low-viscosity discs, we chose to reduce the initial mass to $10\;M_\oplus$.}
\label{table:1}
\end{table}

\subsection{Gas accretion routine} 
\label{section_gasaccroutine}

Gas accretion is modeled using the \cite{Machida2010} and \cite{Kley1999} principles, as in \cite{Crida2017}. The \cite{Kley1999} principle is an arbitrary accretion rate, but it is limited by how much gas is available within the hill sphere of the planet. The \cite{Machida2010} accretion rate is based on 3D isothermal shearing box simulations and represents the accretion rate of the planet in the runaway gas accretion phase. The accretion rate is different than in the contraction phase, as predicted by the core accretion model \citep{Piso2014}.

\cite{Crida2017} used Machida's accretion rate without reducing the gas surface density from its initial value $\Sigma_0$ when computing the accretion rate. To make sure that they were limited to what the disc can provide, they limited the accretion rate to the minimum between Machida's and Kley's accretion rates. As planets create gaps, the surface density profile around the planet is reduced compared to the initial profile. Therefore, in our study, we use the local surface density, defined as the average of the surface density from the position of the planet to $0.9 \; r_H$. Our accretion routine is written as follows:
\begin{equation}
    \dot{M}_p = min
    \begin{cases}
      \dot{M}_{M}=<\Sigma>_{0.9r_H} H^2\Omega \times min\Big[0.14;0.83(r_H/H)^{9/2}\Big]\\
      \dot{M}_{K}= \iint_{S_{disc}} f_{\rm red}(d)\;\Sigma(r,\phi,t)\;\pi d^2f_{\rm acc} \; dr \; d\phi
    \end{cases}      
    \label{eq1}
,\end{equation}
where $S_{disc}$ is the disc surface, $H$ the disc scale height, $\Omega$ the keplerian orbital period of the planet, $d$ the distance from the planet, $ r_H = r_p(m_p/3M_*)^{1/3}$ is the Hill sphere of the planet, and $f_{\rm acc}$ is the inverse timescale upon which the accretion rate of \cite{Kley1999} is occurring. Here, $f_{\rm acc} = 1$ to determine which of the accretion rate is the smallest. $f_{\rm red}$ is a smooth reduction function used to predict what fraction of gas must be accreted on the planet as a function of the distance to the planet. It is defined as:
\begin{equation}
    f_{\rm red} =
    \begin{cases}
        2/3 & \text{if} \; d<0.45 \; r_H \\
        2/3 \times \cos^4\Big(\pi\Big(\frac{d}{r_H}-0.45\Big)\Big) & \text{if} \; 0.45\; r_H<d<0.9 \; r_H
    \end{cases}
    \label{eq2}
.\end{equation}

This function is based on \cite{Robert2018}, where the authors assume that close to the planet gas accretion is $100\%$ efficient ($\rm f_{\rm red} = 1$). However, it seems that $100\%$ efficiency is not realistic in such a case.  \cite{Schulik2019} showed that gas accretion does not start from the full Hill sphere but only from a fraction of it. The accreted mass fraction increases closer to the planet but it does not accrete $100\%$ of the gas in the vicinity. Thus, we chose to limit our study to a maximum  $\rm f_{\rm red}$ value of 2/3.

As Machida's formula only provides information on the amount of mass that should be accreted, we remove the gas in this regime with the same formalism as for the \cite{Kley1999} method. This means that if the planet is accreting in the regime of \cite{Machida2010}, the total amount of gas it accretes is given by Eq.\ref{eq1} and the distribution for where the gas is removed is given by Eq.\ref{eq2}.

This way, the removal scheme of the gas is the same for both regimes, but the mass that can be accreted is either limited by the derived accretion rates of \cite{Machida2010} or by the maximum amount the disc can provide, given by the \cite{Kley1999} method. Unless specified, we remain in the regime where $\rm \dot{M}_M < \dot{M}_K$ throughout, meaning that we always remove the amount of mass suggested in \cite{Machida2010}.

\section{Influence of gas accretion}
\label{section_gasaccinflu}

In this section, we first investigate the difference in gap shape between accreting and non-accreting planets. Then we focus on the influence of different gas accretion rates to explore the variety of gas accretion rates found in the literature. We conclude this section by studying the influence of planetary accretion on stellar accretion.

\subsection{Accretion versus non-accretion} \label{section_accvsnonacc}

\begin{figure}[!t]
        \centering   
        \includegraphics[scale=0.26]{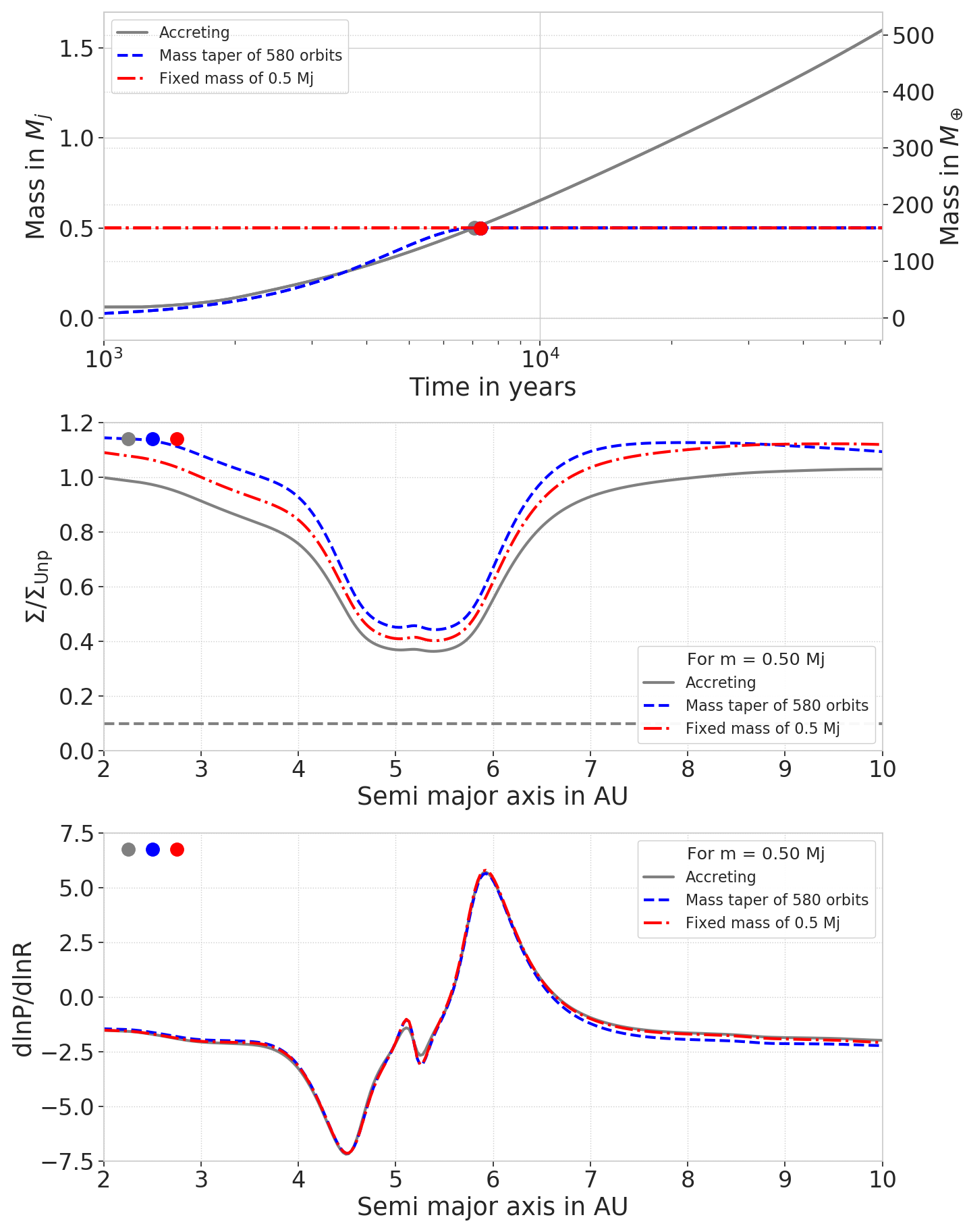}
        \caption{Influence of gas accretion on the gap shape for a planet of m = 0.5 $\rm M_J$, below the nominal gap-opening mass, for $h=0.05$ and $\alpha = 10^{-2}$. \textit{Top:} Time evolution of the planetary mass in the three different simulations: accreting planet (solid gray), mass-tapered planet (dashed blue), and fixed planet mass (dashed red). Mass is removed from the disc only for the accreting example. We plot the mass evolutions after gas accretion is allowed (after t = 100 planetary orbits $\simeq 1.2 \times 10^{3}$ yrs) \textit{Middle:} Perturbed surface density at the time when m = 0.5 $\rm M_J$ (dots in the top panel). The horizontal gray dashed line marks the gap-opening criterion defined by \cite{Crida2006}. An accreting planet will create a deeper gap than a non-accreting one when the gap-opening mass is not yet reached. \textit{Bottom:} Pressure gradient in the three different cases for m=0.5 $\rm M_J$. Similar pressure gradients imply that gas accretion does not have a major influence on the pressure structure in the protoplanetary disc. This implies that particles are trapped at a similar position and we cannot distinguish between accreting and non-accreting planets via the shape of dust traps in this case.}
        \label{accVSnonacc}
\end{figure}

\begin{figure}[!t]
        \centering   
        \includegraphics[scale=0.26]{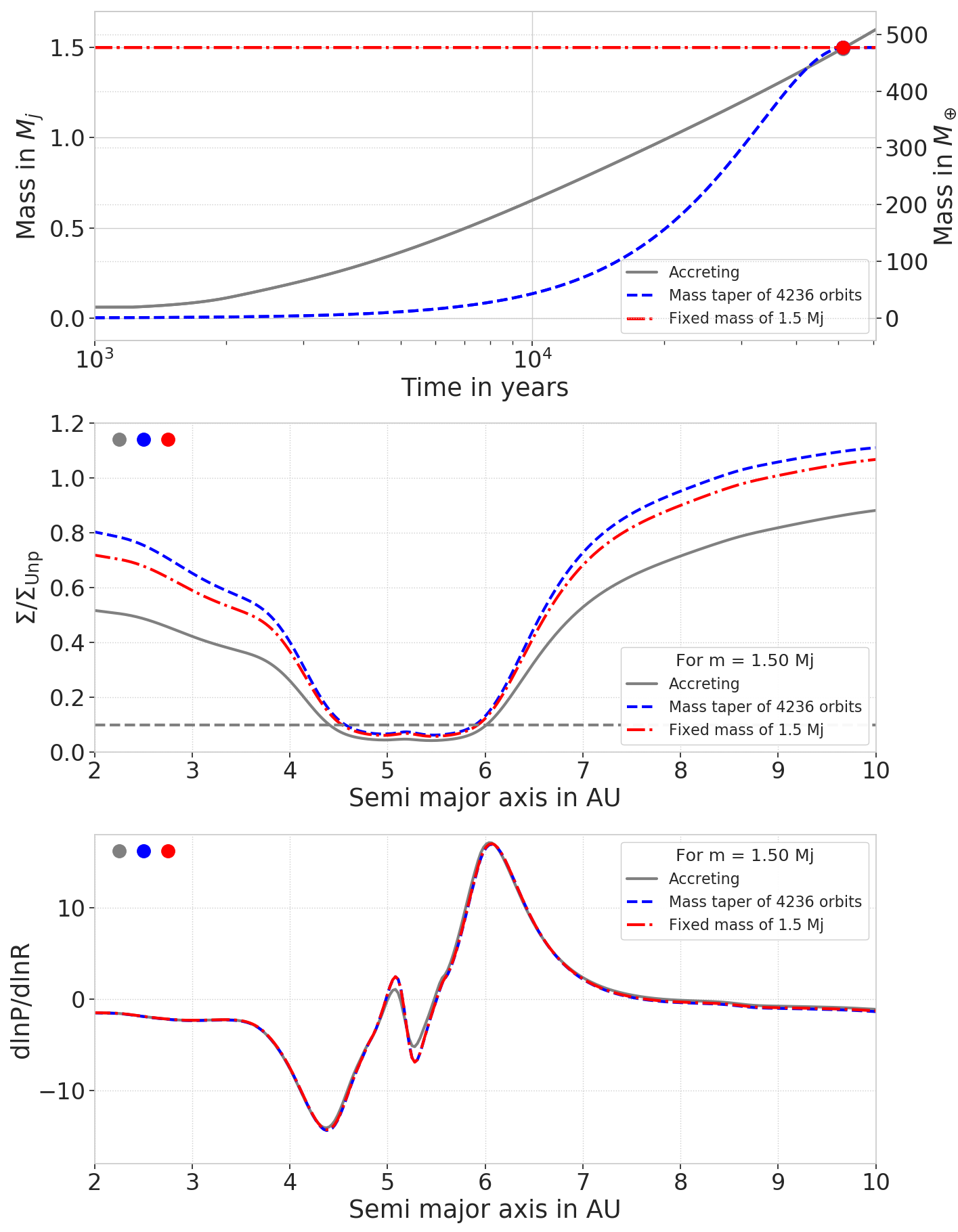}
        \caption{Influence of gas accretion on the gap shape for a planet of m = 1.5 $\rm M_J$, larger than the nominal gap-opening mass, for $h=0.05$ and $\alpha = 10^{-2}$. Panels represent the same quantities as in Fig. \ref{accVSnonacc}. The pressure gradient in the three different cases for m = 1.5 $\rm M_J$ are very similar, as for the 0.5 $\rm M_J$ planets.}
        \label{accVsnonacc1.5}
\end{figure}

In this section, we investigate the effect of gas accretion on the gap shape and the pressure bumps generated by the planet exterior to its orbit. We compare three different simulations: (i) a planet accreting gas following the recipe presented in Sect. \ref{section_gasaccroutine}, where the mass is removed from the disc and added to the planet (solid gray line in the following plots); (ii) a mass-tapered planet where the mass of the planet is changed via the mass-taper function given by Eq. \ref{masstapereq} and no mass is removed from the disc (dashed blue line); and (iii) a planet with its mass fixed to the value where the comparison is made (dotted dashed red line). Each simulation is compared at the moment when the planets reach the same mass $m_{\rm comp}$. For the mass-tapered planet, $n_{\rm orb}$ is chosen so that the planet grows from 0 to $m_{\rm comp}$ over the same number of orbits that the accreting planet needs to reach this mass. We chose to make the comparison before and after reaching the gap-opening mass for $\alpha = 10^{-2}$ and $h = 0.05$, where the gap-opening mass is the mass needed to open a gap of a depth of $\Sigma/\Sigma_{\rm unp} = 0.1$, as in \cite{Crida2006}. We compare the structure of the disc for planets of masses equal to $\rm 0.5\;M_J$ (Fig. \ref{accVSnonacc}) and $\rm 1.5 \; M_J$ (Fig. \ref{accVsnonacc1.5}). Within these parameters, $n_{\rm orb}$ is equal to 580 orbits for $m_{\rm comp} = 0.5 \; \rm M_J$ and to 4236 orbits for $m_{\rm comp} = 1.5 \; \rm M_J$.

On the top panel of Fig. \ref{accVSnonacc}, we show the time evolution of the planetary masses in the three different cases for $\rm m_p$ = 0.5 $\rm M_J$. The mass evolution of the planet growing via the mass taper is very similar to that of the accreting planet. The dots show the mass and time given for the comparison. In the second panel, we show the gap profiles at the time when the accreting planet reaches $\rm 0.5 \; M_J$ (marked by the dots in Fig. \ref{accVSnonacc}). The surface density profile has been normalized to the surface density of a disc without a planet at the same time ($\Sigma_{\rm unp}$) in order to get rid of the effect of natural viscous evolution of the disc. Our simulations show that the accreting planet generates a slightly deeper gap than the planet with a fixed mass, which is in line with the results of \cite{Durmann2015}.

To create a gap, a planet needs to have a strong enough gravitational torque to overcome the pressure torque and viscous spreading that tend to close the gap \citep{Crida2006}. The accreting planet presents a deeper gap as material is removed from the disc as compared to the other two simulations. The difference in surface density at the bottom of the gap is very similar to the difference in surface density in the inner disc due to the fact that mass is removed partially from the inner disc. This will impact the stellar accretion, as we discuss in Sect. \ref{section_ontostar}. For these disc parameters ($\alpha = 10^{-2}, \; h = 0.05$), \cite{Durmann2015} found the same behavior (their Fig. 2). Therefore, relative to the edges of the gap, the gaps have similar depths, but when compared to a disc without a planet, the accreting planet produces a deeper gap, as the non-accreting planets only push material away from their orbit, enhancing the surface density in the inner disc while the accreting planet is removing material. On the other hand, the fixed mass planet is opening a deeper gap than the mass-tapered planet, because the fixed-mass planet exerts a stronger torque on the disc material due to its larger mass up until the time of comparison. At later time, when the mass taper fixes the mass of the planet to $0.5 \; \rm M_J$ and the disc has time to adjust to this planetary mass, the shape of the gaps of the fixed-mass and mass-tapered planets is the same.

At later time, when the planet becomes heavier ($\rm m_p = 1.5 \; M_J$), we can also see on the middle panel of Fig. \ref{accVsnonacc1.5} that the inner disc is more depleted for the accreting planet. Gas accretion plays, therefore, an important role in the depletion of the inner disc. One of the consequences of this behavior is discussed in Sect. \ref{section_ontostar}. The gap depth is only slightly influenced by gas accretion in this case.

The difference in the gap depth can lead to a difference in the torques acting on the planet and, thus, impacting its migration behavior. \cite{Kanagawa2018} found that when the planet creates a deep-enough gap ($\rm \Sigma_{min}/\Sigma_{unp} < 0.6$), then the total torque felt by the planet is proportional to the gap depth. Thus, the difference in gap depth between the accreting and the planet with fixed mass should lead to an equivalent difference in total torque. The difference in gap depth for $\rm m_p = 0.5 \; M_J$ is of about 24$\%$ while the difference in the total torques measured in our simulations is $\sim 17\%$. Using the approach by \cite{Kanagawa2018} regarding the migration behavior and comparing it to our simulations indicate a deviation of $25\%$ between the torques actually measured  and the difference that should occur because of the difference in gap depth. This deviation might lie within the errorbars of the fit derived by \cite{Kanagawa2018}. On the other hand, for the more massive case plotted in Fig. \ref{accVsnonacc1.5}, the difference in gap depth is 38$\%$ and the difference in the total measured torques is $\sim 37\%$. Here, the differences match and our simulations confirm the results from \cite{Kanagawa2018}. Additional simulations with $\rm 1 \; M_J$ planets confirm this trend. Therefore, the differences in torques between an accreting planet and a planet with fixed mass can be explained by their difference in gap depth, as long as the gap is deep enough.

The difference in gap depth and shape will have an influence on the drift of dust particles in the disc. Observations \citep{Teague2018,Andrews2018} often show  rings and gaps in the dust profile of discs, which could be explained by the presence of planets \citep{Pinilla2012,Zhang2018} that generate pressure bumps exterior to their orbits where dust can accumulate \citep{Paardekooper2006_dust}. We show in the lower panels of Figs. \ref{accVSnonacc} and \ref{accVsnonacc1.5} the pressure gradients obtained in the three different cases. As expected from the surface density profiles, the pressure gradients are too similar to display the differences seen through observation. 

In the following sections, we investigate different levels of gas accretion strength and other disc parameters that influence the depth and shape of gaps.

\begin{figure}[!t]
        \centering   
        \includegraphics[scale=0.255]{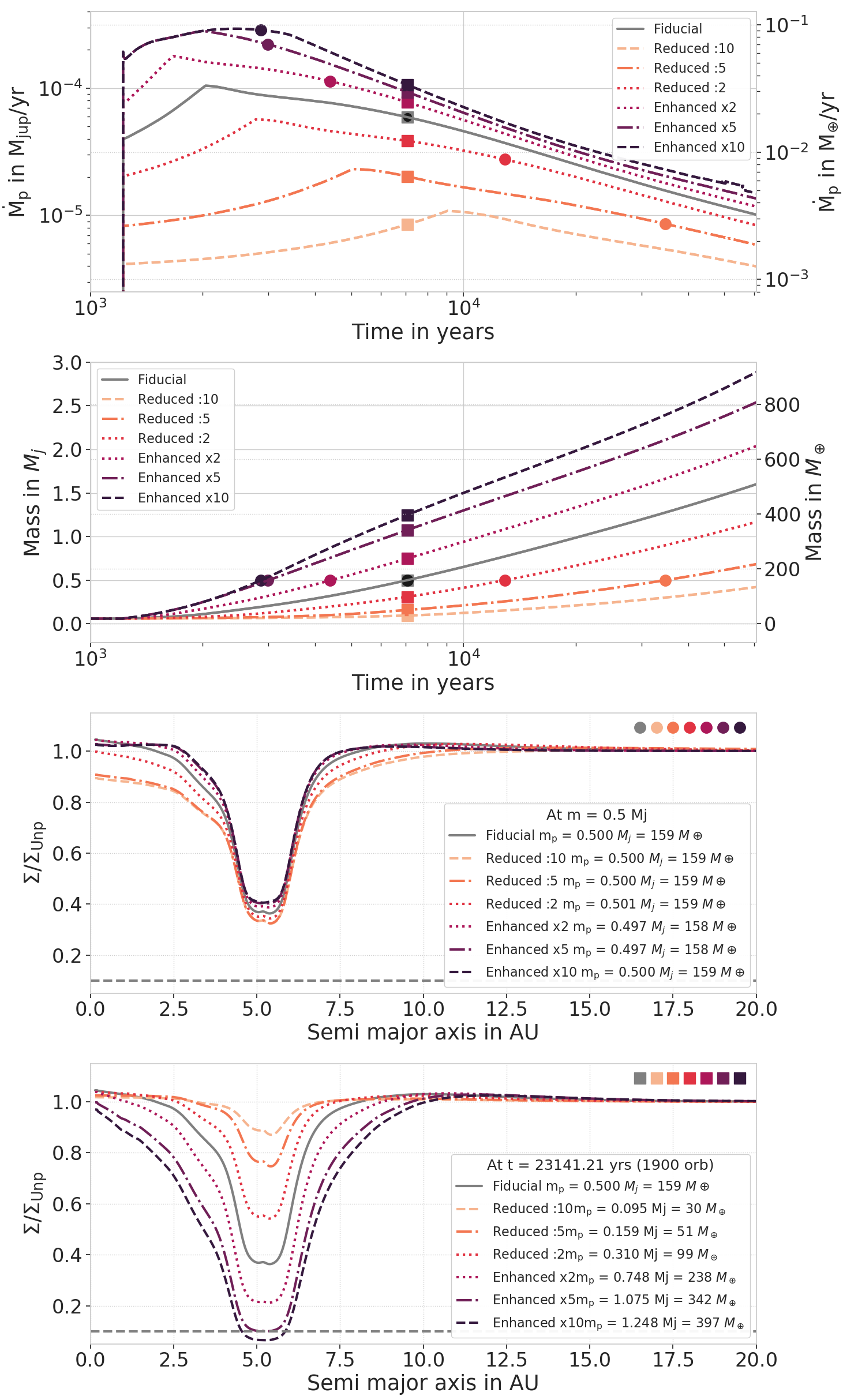}
        \caption{Influence of different accretion rates. \textit{Top:} Time evolution of the different accretion rates. The fiducial setup is represented in solid gray line. The dashed, dotted, and dotted-dashed lines represent the different accretion rates applied here, where the Machida part was divided and multiplied by 10, 5, and 2. The enhanced by 10 and 5 simulations show a similar rate at the beginning of the simulation as they are limited to the maximum accretion the disc can provide. \textit{Middle 1:} Time evolution of the planetary mass in the seven different accretion regimes. The dots represent the time at which the mass is 0.5 $\rm M_J$ and the squares represent the different masses at the same time. \textit{Middle 2:} Perturbed surface density at the time where m = 0.5 $\rm M_J$ (dots in previous panels). The surface density is normalized to a disc without a planet. \textit{Bottom:} Perturbed surface density at the same time t = 580 orbits (squares in previous panels). The horizontal gray dashed line marks the gap-opening criterion as defined by \cite{Crida2006},$\Sigma/\Sigma_{\rm unp} = 0.1$.}
        \label{diffacc}
\end{figure}

\subsection{Influence of different gas accretion rates} \label{section_diffacc}

Gas accretion rates onto planets are not very precisely constrained. In the runaway gas-accretion phase accretion, rates range from $\rm \dot{M}_p \simeq 10^{-7} \; M_{J}/yr \simeq 3 \times 10^{-5} \; M_\oplus/yr$ to $\rm \dot{M}_p \simeq 10^{-4} \; M_{J}/yr \simeq 3 \times 10^{-2} \; M_\oplus/yr$ \citep{DAngelo2003,Ayliffe2009,Machida2010,Schulik2019}. Recent studies claim that the accretion rates could be even smaller; \cite{Lambrechts2019} found that the mass flux through the envelope is different from the gas accretion rate, therefore, even if the mass flux is on the order of $\rm 10^{-5} \; M_{J}/yr$, the gas accretion rate is actually 10 to 100 times smaller. \cite{Tanigawa2016} derived an even smaller accretion rate when accounting for the gap-opening ($\rm \dot{M}_p \simeq 10^{-8} \; M_{J}/yr \simeq 3\times10^{-6} \; M_\oplus/yr$). In order to reflect this discrepancy, we investigate different accretion rates in this subsection by scaling our nominal Machida accretion rates (Eq.\ref{eq1}) by a factor of between 0.1 to 10. 

In the top panel of Fig. \ref{diffacc}, we show the seven different accretion rates obtained when we scale the Machida accretion rate by a factor of 0.1, 0.2, 0.5, 1, 2, 5, and 10. The darker color represents a higher accretion rate. The maximum of the accretion rate is shifted to later time if the accretion rate is reduced. This maximum is reached when the accretion rate (Eq.\ref{eq1}) switches from a regime dominated by the Hill sphere to a regime proportional to 0.14. The critical mass that is needed for this switch in accretion rate is reached at later times for reduced accretion rates. 

As gas accretion is limited by what the disc can provide, the accretion rates in the case of an enhancement by 5 and 10 are the same at the beginning of the simulation: they start in the Kley accretion regime (eq. \ref{eq1}). Both simulations switch to the Machida accretion regime and become different after a certain time. This indicates that most of the time the planet is accreting less than what the disc can provide. The values of the accretion rates range from $\rm \sim 6\times10^{-6} \; M_J/yr$ to $\rm \sim 2\times10^{-4} \; M_J/yr$, corresponding to the different values reported in the literature \citep{DAngelo2003,Ayliffe2009,Machida2010,Tanigawa2016,Schulik2019,Lambrechts2019}. We note, however, that the accretion rates are reduced over time, where at later stages, our measured accretion rates are below $\rm 2 \times 10^{-5} \; M_J/yr$. This indicates that the gap-opening process of planets plays a role in setting the accretion rate.

In the second panel, we show the time evolution of the planetary mass in the seven different cases. The biggest planet formed after $6\times10^{4}$ years is slightly less than $\rm 3 \; M_J$, obviously, this is the case of the highest accretion rate. In addition to the assumptions of our simulation setup, time evolution plays a crucial role. A longer integration time would lead to a larger planetary mass. In reality, a planet would also migrate to the inner disc, where the gas surface density is larger and thus the accretion rate could be higher, even if the Hill sphere of the planet shrinks while getting closer to the central star, depending on the surface density profile slope \citep{Crida2017}.

\begin{figure}[!t]
        \centering   
        \includegraphics[scale=0.25]{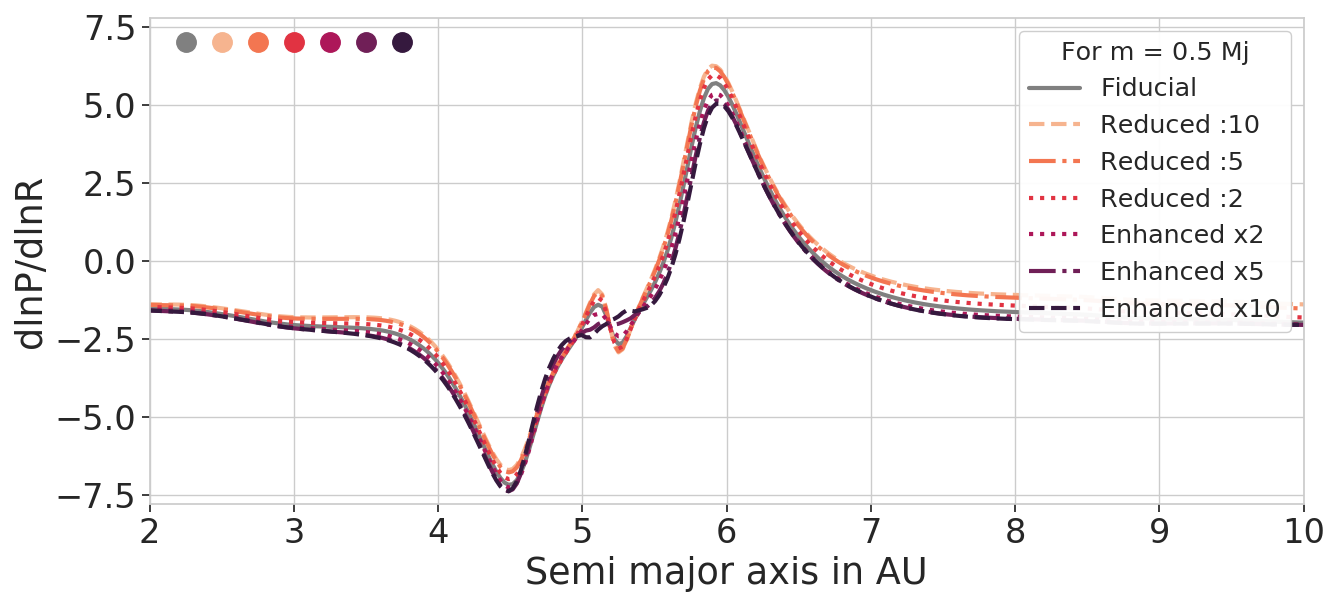}
        \caption{Pressure gradient in different gas accretion models for m = 0.5 $\rm M_J$. Even though the different models present different inner disc structures at the same planetary mass, the difference in the pressure gradients is too small to be observationally resolved. The bump located at the planet position is due to the material around the planet location. It vanishes with higher accretion rates as the planet accretes all the material at this location.}
        \label{pressurebump_diffacc}
\end{figure}

In the third and fourth panels of Fig. \ref{diffacc}, we show the azimuthally averaged surface density profiles. We normalized the profiles to the profile of a disc without a planet as explained in Sect. \ref{section_accvsnonacc}. This way we compare the actual shape of the gap. For this viscosity and aspect ratio, $\alpha = 10^{-2}$ and $h = 0.05$, gas accretion seems to have only a small effect on the gap shape, resulting in a slightly deeper gap for the most reduced accretion rate. The difference in gap depth is due to the depletion of the inner disc: for the reduced accretion rate, the planet reaches $0.5 \; \rm M_J$ at a later time, giving time to the inner disc to be depleted via viscous spreading towards the star. We discuss this in more detail in the next section. As the gap shapes are only slightly influenced by gas accretion, one can expect the pressure bumps to be similar, which is shown in Fig. \ref{pressurebump_diffacc}. The difference in the disc pressure profiles for planets with different accretion rates are very small, making it indistinguishable via observation, in contrast to migrating planets \citep{Meru2019, Weber2019}. However, in the next section, we investigate the influence of planetary gas accretion on another observable: gas accretion onto the star.

\subsection{Influence on the stellar accretion rate} \label{section_ontostar}

\begin{figure}[t]
        \centering
        \includegraphics[scale=0.23]{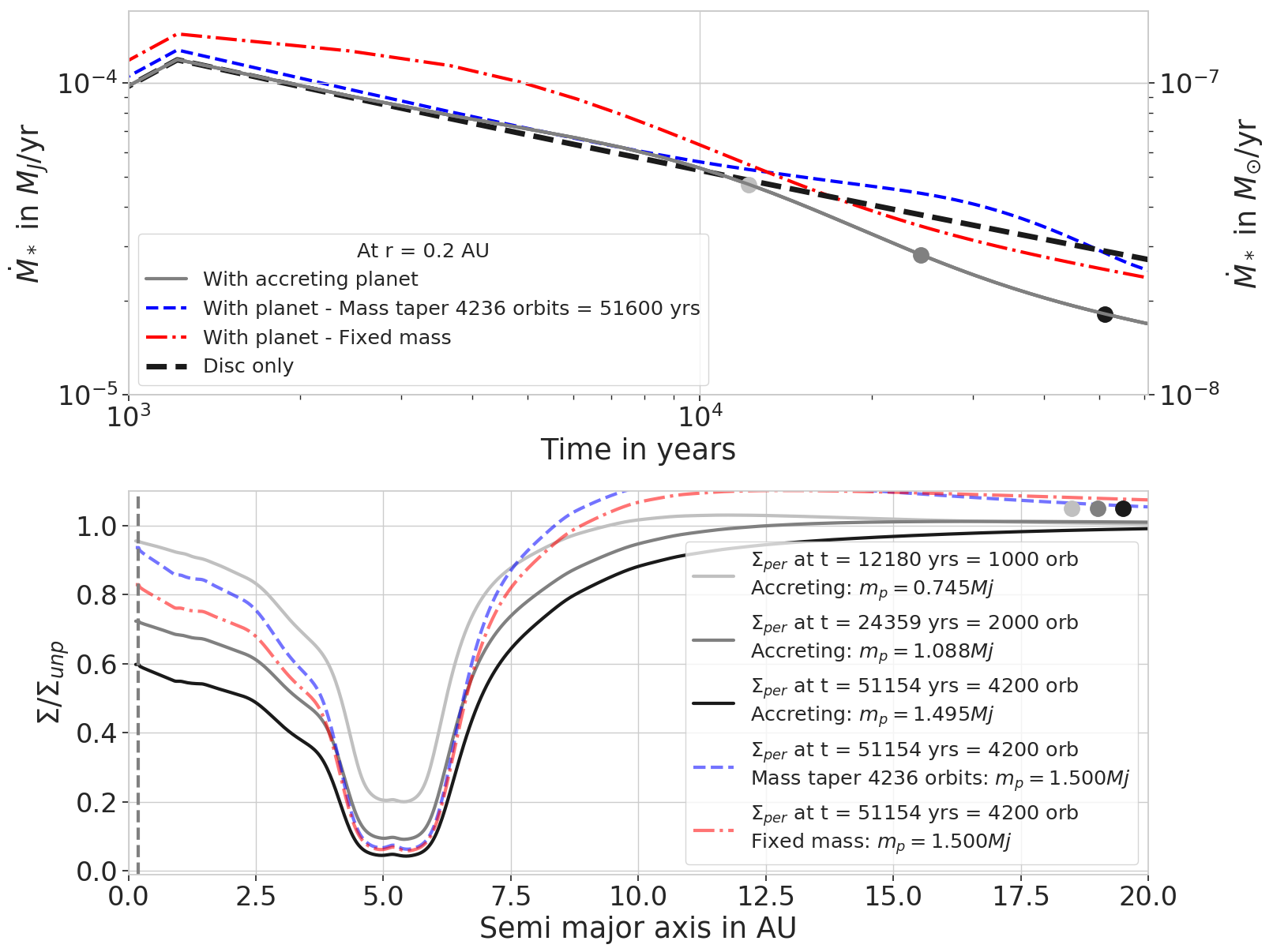}
        \caption{Influence of planetary gas accretion on the stellar gas accretion for our fiducial disc parameters ($h = 0.05, \; \alpha = 10^{-2}$). \textit{Top:} Time evolution of the stellar gas accretion rates at the inner disc (0.2 AU). The different lines represent different simulations: with an accreting planet (solid gray), with a mass-tapered planet (dashed blue), with a fixed-mass planet (red dotted dashed). and without a planet (bold black dashed). We can see here that an accreting planet decreases the stellar accretion rate. \textit{Bottom panel:} Perturbed surface density at three different times for the accreting case and after 4200 orbits in the mass tapered and fixed mass cases. The dashed vertical line is located at 0.2 AU, where we measure the stellar accretion rate. The three different times correspond to the dots in the upper panel. The time evolution of the perturbed gas surface density shows how the inner disc is slowly depleted by the viscous stellar accretion but also by the accreting planet. In comparison, the mass tapered and fixed mass cases present a less depleted inner disc, due to the absence of planetary gas accretion (see also Fig. \ref{accVsnonacc1.5}).}
        \label{ontostarVSplanet}
\end{figure}

Gas accretion onto the star is an observable feature that can be measured via the UV excess in a star SED \citep{Hartmann1994,Calvet1998}. Here, we study the influence that a gas-accreting planet has on stellar gas accretion compared to a disc without a planet. We define the stellar gas accretion rate as the mass flux through the inner boundary of the disc: $\rm \dot{M}_* = -2\pi \; r_{in} v_{r,in} \Sigma_{in},$ where $\rm r_{in} = 0.2$ AU and $\rm v_{r,in}$ and $\rm \Sigma_{in}$ are the radial velocity and surface density at $\rm r_{in}$, respectively. We compare in Fig. \ref{ontostarVSplanet}, the stellar gas accretion of a disc with an accreting planet (solid gray line), a fixed-mass planet (dotted dashed red), a mass-tapered planet (dashed blue), and a disc without a planet (bold dashed black). The two non-accreting planets have a mass of $\rm m_p = 1.5 \; \rm M_J$, as in Fig. \ref{accVsnonacc1.5}, leading to a mass taper of $n_{\rm orb} = 4236 \rm \; orbits \simeq 51 600 \; yrs$.

\begin{figure}[t]
        \centering   
        \includegraphics[scale=0.23]{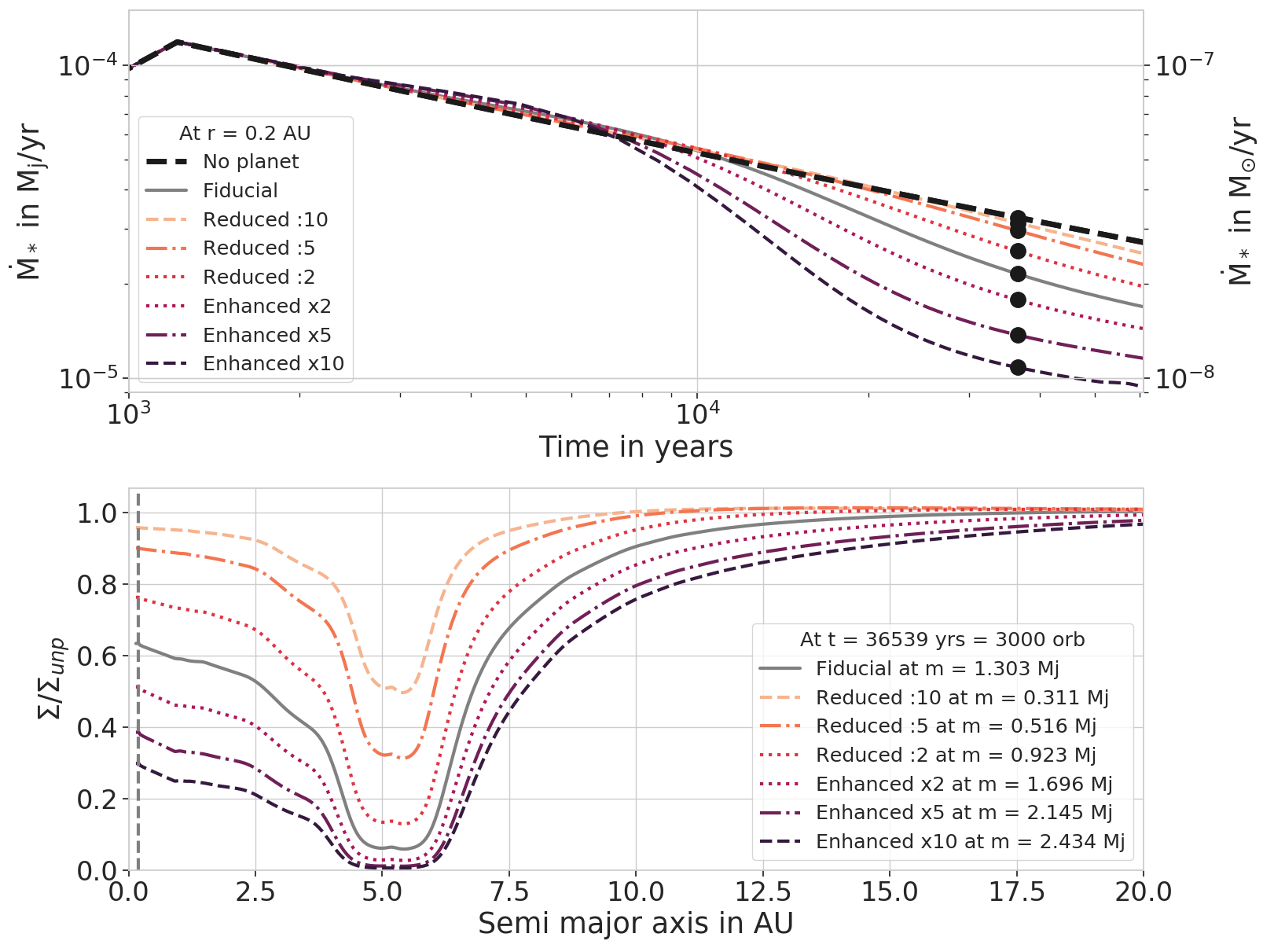}
        \caption{Influence of different planetary gas accretion rates on the stellar gas accretion for the fiducial disc parameters ($h = 0.05, \; \alpha = 10^{-2}$). \textit{Top:} Time evolution of the accretion rates onto the star at the inner disc (0.2 AU) for eight different cases: the seven different planetary accretions rates (enhanced, fiducial, and reduced) and a disc without a planet (bold black dashed line). A more efficiently accreting planet decreases the stellar accretion rate more efficiently. This is due to the depletion of the inner disc induced by the planetary accretion. The black dots mark the time at which the surface density profiles are plotted. \textit{Bottom panel:} Perturbed surface density at time t = 3000 orbits = 36 539 years for the seven different cases. The vertical gray dashed line lies at 0.2 AU which is the location of the inner disc, where the stellar accretion is measured.}
        \label{ontostardiffacc}
\end{figure}

At the start of the simulations, only the fixed-mass planet shows a significantly higher stellar accretion rate than the others. This is due to the fact that a massive planet of 1.5 $\rm M_J$ was introduced into the disc in only three orbits. Therefore, the disc is highly perturbed and the planet pushes a significant quantity of material in the inner disc. As $\rm \dot{M}_* \propto \Sigma$, the stellar gas accretion is then enhanced. At the same time, the mass-tapered and accreting planets are slowly growing into the disc, pushing less material to the inner disc at the beginning of the simulation. 

Even though \cite{Rafikov2016} showed that spiral arms can drive accretion via depositing angular momentum, these spiral arms are too weak at this location to make a difference. They also showed that such accretion may occur at the end of the lifetime of the disc, as it corresponds to very short accretion timescales. In consequence, the stellar accretion rates for the mass-tapered planet, the fixed-mass planet, and the disc without a planet are all similar. 

On the other hand, we see that after $\sim 10^4$ years, the accretion rate differs from one simulation to another. First, the fixed-mass planet shows an accretion rate that is slightly lower than the disc without a planet. Even though the gap might prevent some material from reaching the inner disc, there is still a large flow of material through the gap to keep the inner disc from becoming significantly depleted. This can be seen in the middle panel of Fig. \ref{accVsnonacc1.5}, where the surface density in the inner disc of the fixed mass case is only slightly affected by the presence of the planet. This is due to the high viscosity of this simulations ($\alpha = 10^{-2}$). Before the end of the simulation, the stellar accretion rate in the case of the mass-tapered planet becomes slightly higher than the accretion of the disc without a planet. In the top panel of Fig. \ref{accVsnonacc1.5}, we see that the mass taper function makes the planet grow rapidly after $\sim 10^4$ years, resulting in the same effect as for the fixed-mass planet: as the planet grows rapidly, it pushes a large quantity of material into the inner disc, enhancing the stellar accretion rate, and as the gap profile tends to reach an equilibrium at this stage, it will reach the same flow through the gap as the fixed mass planet.

On the other hand, the presence of the accreting planet decreases the stellar accretion rate. This is due to the fact that the planet, besides accreting material from the outer disc and therefore preventing a large amount of gas to flow through the disc, accretes gas from the inner disc as well. This helps deplete the inner disc quicker than through the viscous accretion naturally present in the disc. The depletion of the inner disc by planetary gas accretion can be seen in the lower panel of Fig. \ref{ontostarVSplanet}, with the surface density profiles of the accreting planet shown at different times: $\rm \Sigma/\Sigma_{unp} < 1$, as $\rm \Sigma_{unp}$ is defined to be the surface density of the empty disc. As in Figs. \ref{accVSnonacc} and \ref{accVsnonacc1.5} (middle panels), we also see  that the depletion is enhanced by the accretion onto the planet compared to the two non-accreting planets.

However, we notice that the accretion rate onto the planet can be higher than the stellar accretion rate (top panels of Figs. \ref{diffacc} and \ref{ontostardiffacc}). The planet can accrete more than what the disc can supply only at the very beginning of the simulation as it is emptying its horseshoe region. Then it accretes material from the inner and outer parts of its orbit: this is where planetary accretion is limited to what the disc can provide. The reason the stellar gas accretion rate is smaller than the planetary accretion rate after a longer time is that the stellar accretion rate is measured at the inner edge of the disc (mass flux at r = 0.2 AU), whereas the planet is accreting material from the inner and the outer disc (inside and outside r = 5.2 AU). We compared the stellar flux at a radius located beside the planet (i.e., at 7.5 AU while the planets is located at 5.2 AU) and we observed that the planet's accretion rate becomes limited by the mass flux at this location after a certain time: the time needed to empty its horseshoe region and a large part of the inner disc. In conclusion, planetary accretion is limited to what the outer disc can provide as the inner disc is depleted by the planet and the viscous accretion towards the star.

To investigate the influence of planetary gas accretion into the depletion of the inner disc, we show in Fig. \ref{ontostardiffacc}, the stellar accretion rate for all our enhanced and reduced planetary accretion rates. The top panel shows that larger planetary accretion rates result in lower stellar accretion rates. As the planet accretes more gas, it depletes the inner disc faster, reducing the stellar accretion rate. At $\rm t = 36500 \; yrs = 3000$ orbits, the time shown with the black dots on the top panel, the lowest stellar accretion rate is roughly three times lower than the highest rate. However, the corresponding planetary accretion rates differ by a factor of four (Fig. \ref{diffacc}). This means that the stellar accretion does not directly scale with the planetary accretion. This is due to the fact that the depletion of the inner disc also highly depends on the viscosity of the disc. Here, the viscosity is high: with $\alpha = 10^{-2}$, planetary gas accretion and viscous spreading through the gap act together to deplete the inner disc. Therefore, if the planetary gas accretion rate is not high enough, then the inner disc is replenished by viscously spreading gas, whereas when the planetary accretion rate is high, the inner disc does not have time to be replenished by viscous spreading. In Sect. \ref{section_diffvisco}, we study the influence of the viscosity on the stellar accretion and discuss its consequences in Sect. \ref{section_discussion_accontostar}. 

In conclusion, at high viscosity, the planetary gas accretion rate has a strong influence on stellar accretion. In the following section, we investigate the influence of two disc parameters: the aspect ratio and the viscosity, which can have an influence on the accretion onto the planet and onto the star.

\section{Influence of disc parameters} \label{section_influenceofparams}

Disc parameters have a strong impact on how a planet carves a gap \citep{LinPapaloizou1986,Crida2006,Fung2014,Kanagawa2015}. Therefore, we can expect them to have a strong influence on gas accretion as well. Here, we study the influence of the aspect ratio and the viscosity on the planetary gas accretion rate before studying the influence on the gap-opening mass. Finally, as migration depends on the disc parameters too, we investigate how planetary gas accretion can alter the migration of outer planets in the disc by studying migration maps.

\subsection{Different aspect ratios} \label{section_diffh}

\begin{figure}[!t]
        \centering   
        \includegraphics[scale=0.24]{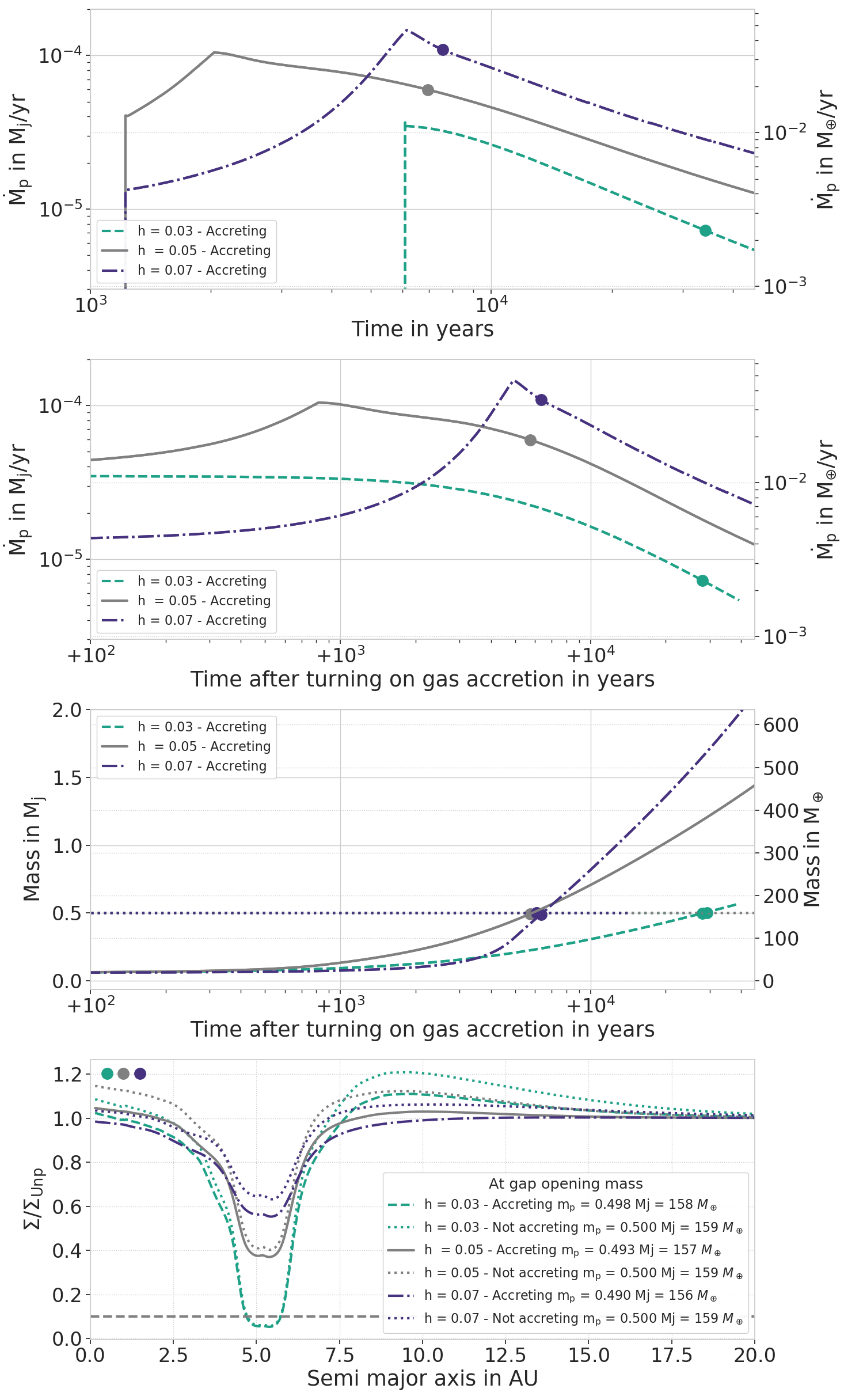}
        \caption{Influence of different aspect ratios with $\alpha = 10^{-2}$ and $\rm m_{init} = 20\;M_\oplus$. \textit{Top:} Time evolution of the planetary accretion rate. The flip in the accretion rate occurs at different moment (i.e., different planet mass) due to the dependence on $h$ of the Machida accretion rate (Eq.\ref{eq1}). As the time needed for the initial mass to reach an equilibrium is higher for lower aspect ratios, gas accretion starts later for $h = 0.03$ (see Table \ref{table:2}). \textit{Middle 1:} Evolution of the accretion rates as a function of the time after gas accretion is turned on. \textit{Middle 2:} Time evolution of the planetary mass. Lower aspect ratios result in more massive planets. \textit{Bottom:} Perturbed surface density at the time where $\rm m_p = 0.5 \; M_J$ (dots on previous panels). The two effects added up here make the gap shapes really different, namely, the influence of the aspect ratio on the shape of the gap and the time at which the planet reaches $\rm 0.5 \; M_J$. We compare the gap shapes of accreting planets with gaps created by fixed mass planets: depending on the aspect ratio, gas accretion has a different influence. The horizontal gray dashed line marks the gap-opening criterion as defined by \cite{Crida2006}.}
        \label{diffaspectratio}
\end{figure}

The aspect ratio of the disc impacts accretion in two ways. The first is related to gas accretion (Eq. \ref{eq1}), which is, to some extent, proportional to $(r_H/H)^{9/2}$. At the beginning of the simulation, the planet remains in this regime until the Hill sphere reaches 2/3 of the disc scale height and then switches to the regime where $\dot{M} \propto 0.14$. In \cite{Machida2010}, the different regimes are explained by the switch from a regime regulated by Bondi accretion to a regime governed by Hill accretion. This switch is visible in our work in the time evolution of the gas accretion rates in Figs. \ref{diffacc}, \ref{ontostarVSplanet}, \ref{diffaspectratio}, and \ref{differentvisco}. The aspect ratio has a direct influence on the mass at which the switch occurs. In top panels of Fig. \ref{diffaspectratio}, we show the resulting planetary accretion rates for different aspect ratios, with $h = 0.03, 0.05$ and $0.07$. As the aspect ratio increases, the switch occurs at higher masses and therefore at later time. For $h = 0.03$, the simulation directly starts in the regime $\propto 0.14$ (Eq. \ref{eq1}) as the mass at which $r_H \simeq 2H/3$ is smaller than our initial mass of $\rm 20 \; M_\oplus$ ($\rm m_{switch} = 7.8 \; M_\oplus $). 

We show in the lower panel of Fig. \ref{diffaspectratio}, the perturbed gas surface density profiles for the different aspect ratios and compare accreting planets with planets that have a fixed planetary mass of $\rm 0.5 \; M_J$. Gas accretion has a different impact on the gas surface density profile depending on the aspect ratio. For $h = 0.07$, the gap is deeper with accretion than without accretion. For $h = 0.05$, we find back the same result from Sect. \ref{section_accvsnonacc}, where gas accretion has only a slight impact on the gap depth. Finally, for $h = 0.03$, the planet has reached its gap-opening mass, which shows a similar-looking for both the accreting and non-accreting cases. Comparing a planet with a fixed mass of $\rm 0.2\;M_J$ for $h=0.03$ (below the gap opening mass defined by \citealt{Crida2006}) with an accreting planet of the same mass (not shown) reveals a deeper gap in the case of the fixed mass planet. Our results show three different behaviors depending on the aspect ratio: 1) gas accretion helps creating a deeper gap (large h); 2) it has almost no impact ($h=0.05$); or 3) it prevents the creation of a deeper gap (low h). These three behaviors are explained in Sect. \ref{section_gapopeningmass}, where we explore the impact of gas accretion on the gap-opening mass. In the case of a low aspect ratio, one should note that a $\rm 20\;M_\oplus$ mass planet already creates a deep gap at low viscosities. We chose to keep this initial mass in this section in order to compare it with the higher aspect ratios as the comparison is made at high viscosity. However, we reduce the initial mass down to $\rm 10\;M_\oplus$ in Sect. \ref{section_gapopeningmass} to study the impact on the gap-opening mass as a function of the disc parameters.

As the accretion rates are very different from one aspect ratio to the other, the planet masses are diverse as well. The middle panel of Fig. \ref{diffaspectratio} shows that a later switch in accretion rates results in more massive planets. As a result, our simulations indicate that planetary gas accretion is more efficient in hotter discs. As it is harder to create a deep gap in hotter discs, the surface density at the location of the planet is larger (lower panel of figure \ref{diffaspectratio}) and allows for a more efficient planetary gas accretion. On the other hand, if the disc is hotter, it is harder for a planet to reach the pebble isolation mass \citep{Lambrechts2014,Bitsch2018,Ataiee2018}, where the planet generates a small pressure bump exterior to its orbit preventing solid accretion. In addition, if the disc is hotter, it is more difficult to accrete pebbles efficiently because they are less concentrated in the disc's midplane \citep{Youdin2007}, resulting in a less efficient formation of planetary cores \citep{Ndugu2018}. Therefore, it is easier to form giant planets via runaway gas accretion when their core is already formed but it is harder to initially form these cores in hotter discs.

\begin{table}            
\centering                          
\begin{tabular}{c c c c}        
\hline\hline                 
Parameters & $\alpha$ & $\Sigma/\Sigma_{\rm unp}$ & $\rm t_{gap,init}$ in orbits \\    
\hline\hline                      
    $h = 0.03$ & $10^{-2}$ & 0.7414 & 500\\
    $\rm m_{init} = 20\;M_\oplus$ &  &  & \\
    \hline
     & $2\times10^{-2}$ & 0.9823 & 100 \\     
   $h = 0.05$ & $10^{-2}$ & 0.9678 & 100 \\ 
   $\rm m_{init} = 20\;M_\oplus$ & $10^{-3}$ & 0.7317 & 900 \\ 
     & $10^{-4}$ & 0.2949 & 3300 \\ 
   \hline
     & $10^{-2}$ & 0.9953 & 100 \\     
   $h = 0.07$ & $5\times10^{-3}$ & 0.9923 & 100 \\ 
   $\rm m_{init} = 20\;M_\oplus$ & $10^{-3}$ & 0.9644 & 300 \\ 
     & $10^{-4}$ & 0.7795 & 1700 \\ 
   \hline
   $h = 0.03$ & $10^{-2}$ & 0.8912 & 300 \\ 
   $\rm m_{init} = 10\;M_\oplus$ & $10^{-3}$ & 0.4859 & 1300 \\ 
    & $5\times10^{-4}$ & 0.3408 & 1900 \\
\hline                                   
\end{tabular}
\caption{Depth of the initial gap and time needed to reach the equilibrium as a function of the disc parameters. Lower viscosities ($\nu = \alpha h R c_s$) imply deeper initial gap and larger gap opening times. For $h = 0.03,$ we used two different initial masses: $\rm m_{init} = 20\;M_\oplus$ in Sect. \ref{section_diffh} to compare with higher aspect ratios but $\rm m_{init} = 10\;M_\oplus$ in Sect. \ref{section_gapopeningmass} as a $20 M_\oplus$ already opens a deep gap at low viscosities.}
\label{table:2}
\end{table}

\subsection{Different viscosities} \label{section_diffvisco}

\begin{figure}[t]
   \centering   
   \includegraphics[scale=0.25]{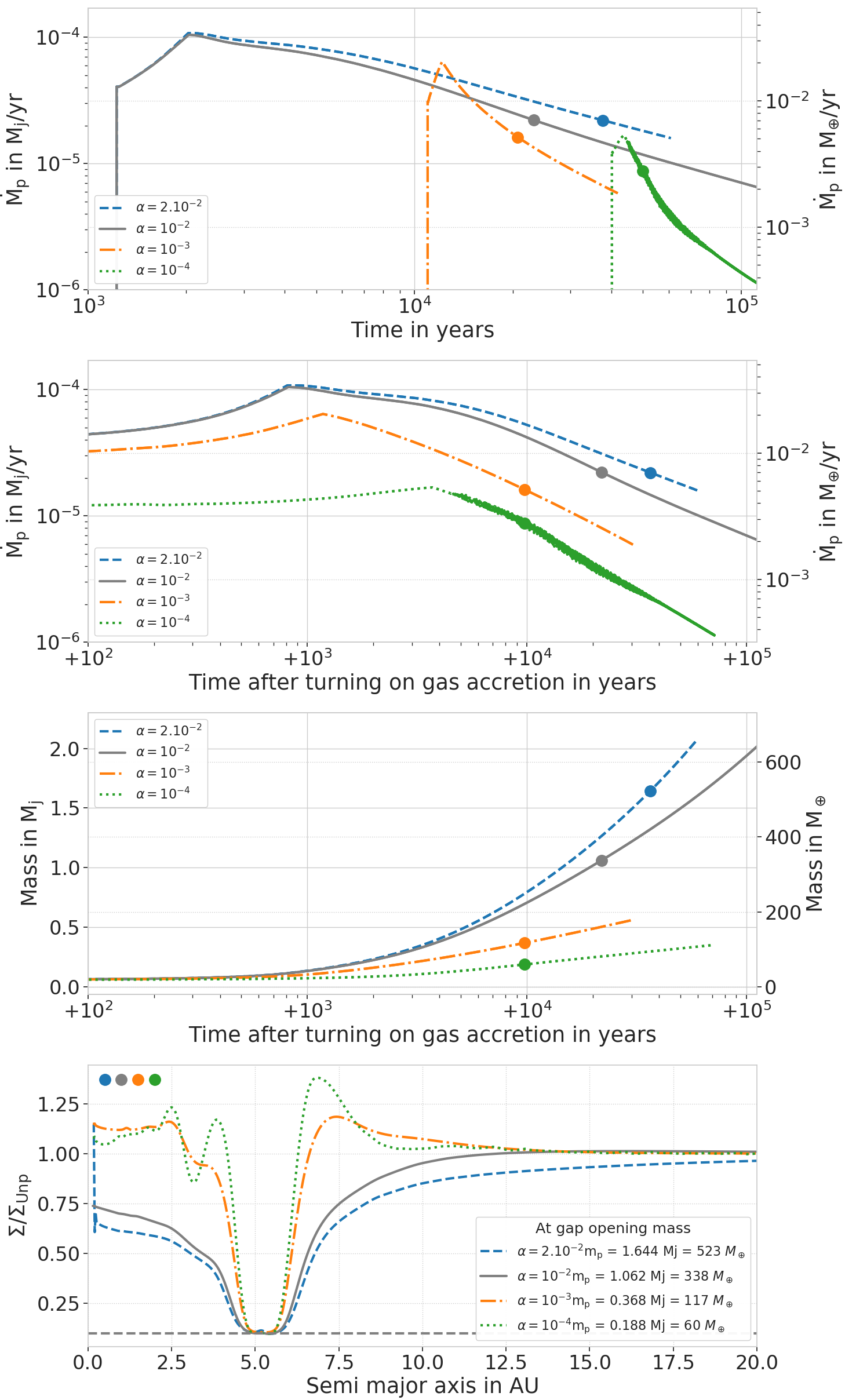}
   \caption{Influence of different viscosities for $h=0.05$. \textit{Top:} Time evolution of the accretion rate onto the planet. As the time needed for the initial mass to reach an equilibrium is highly dependent on the viscosity, gas accretion starts later for lower viscosities. \textit{Middle 1:} Evolution of the accretion rates as a function of the time after gas accretion is turned on. As the viscosity is lowered, the Rossby Wave Instability is triggered: for $\alpha \lesssim 10^{-4}$, vortices are formed and influence the planetary gas accretion rates, explaining the oscillations in the accretion rate curves. \textit{Middle 2:} Evolution of the planetary mass as a function of the time after gas accretion is turned on. The dots represent the time at which the gap-opening mass is reached. \textit{Bottom:} Perturbed surface density at the time where gap-opening mass is reached (dot on the other panels). The gap-opening mass is defined by the mass needed to reach $\Sigma/\Sigma_{\rm unp} = 0.1$ \citep{Crida2006}. It is represented here by the horizontal gray dashed line.}
   \label{differentvisco}
\end{figure}

\begin{figure}[!t]
   \centering   
        \includegraphics[scale=0.23]{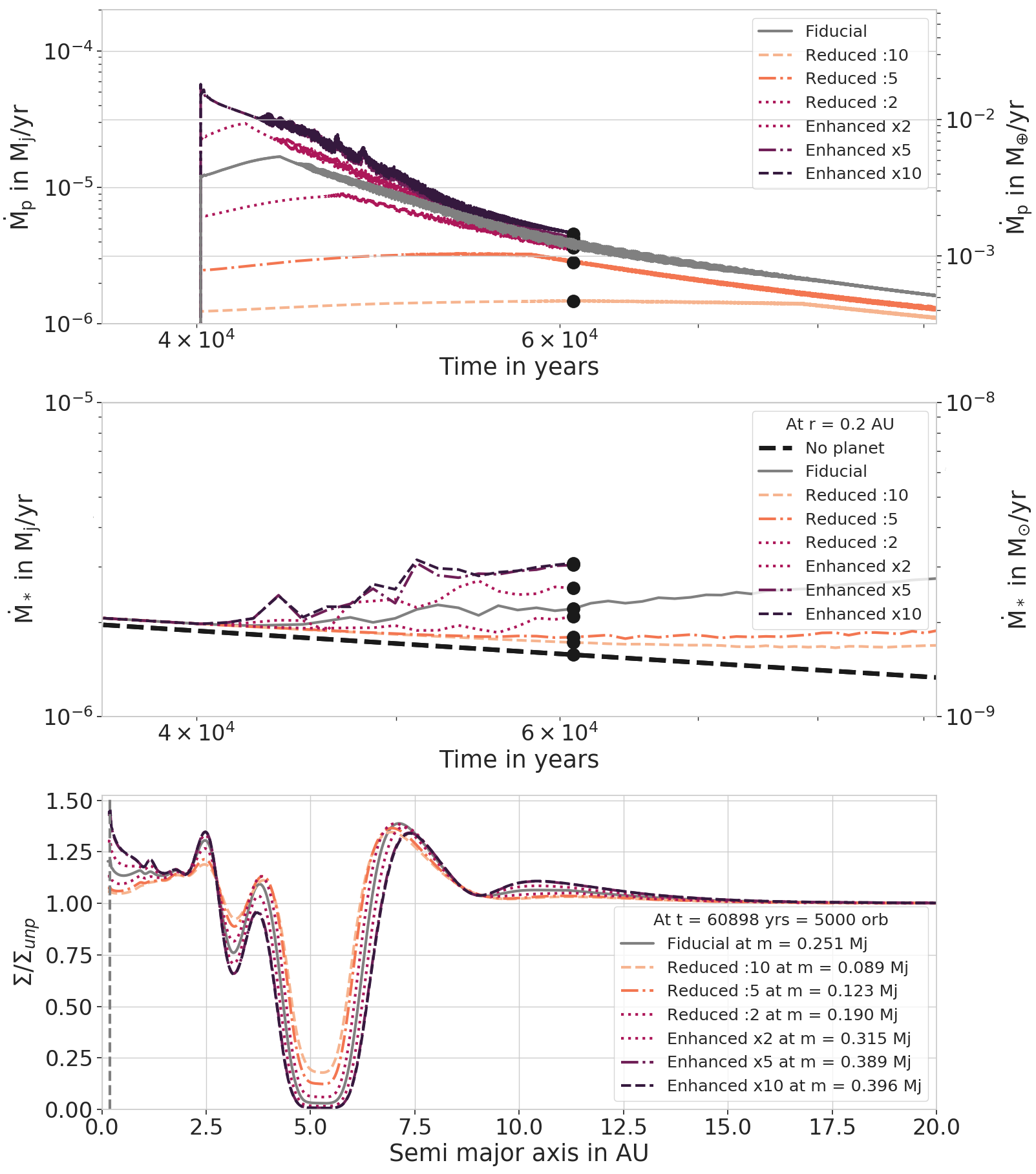}
        \caption{Influence of different accretion rates on stellar gas accretion for $\alpha = 10^{-4}$ and $h=0.05$. \textit{Top:} Planetary gas accretion rate for the seven reduced, enhanced, and fiducial accretion rates. The presences of vortices create oscillations in the accretion rate: the larger the vortex, the larger the oscillations. \textit{Middle:} Stellar gas accretion at the inner edge of the disc (0.2 AU). At this low viscosity, the trend is flipped compared to the high viscosity case: a more efficiently accreting planet will lead to an enhanced accretion onto the star. This is due to the fact that at low viscosity, the inner disc is mostly perturbed by the presence of the planet and the viscosity is too low to compensate for it. \textit{Bottom:} Perturbed surface density at time t = 4000 orbits = 48 718 yrs. The vertical line shows the location at which the stellar accretion rates are measured (0.2 AU).}
        \label{ontostardiffacc_alpha4}
\end{figure}

\begin{figure*}[!ht]
   \centering   
        \includegraphics[scale=0.29]{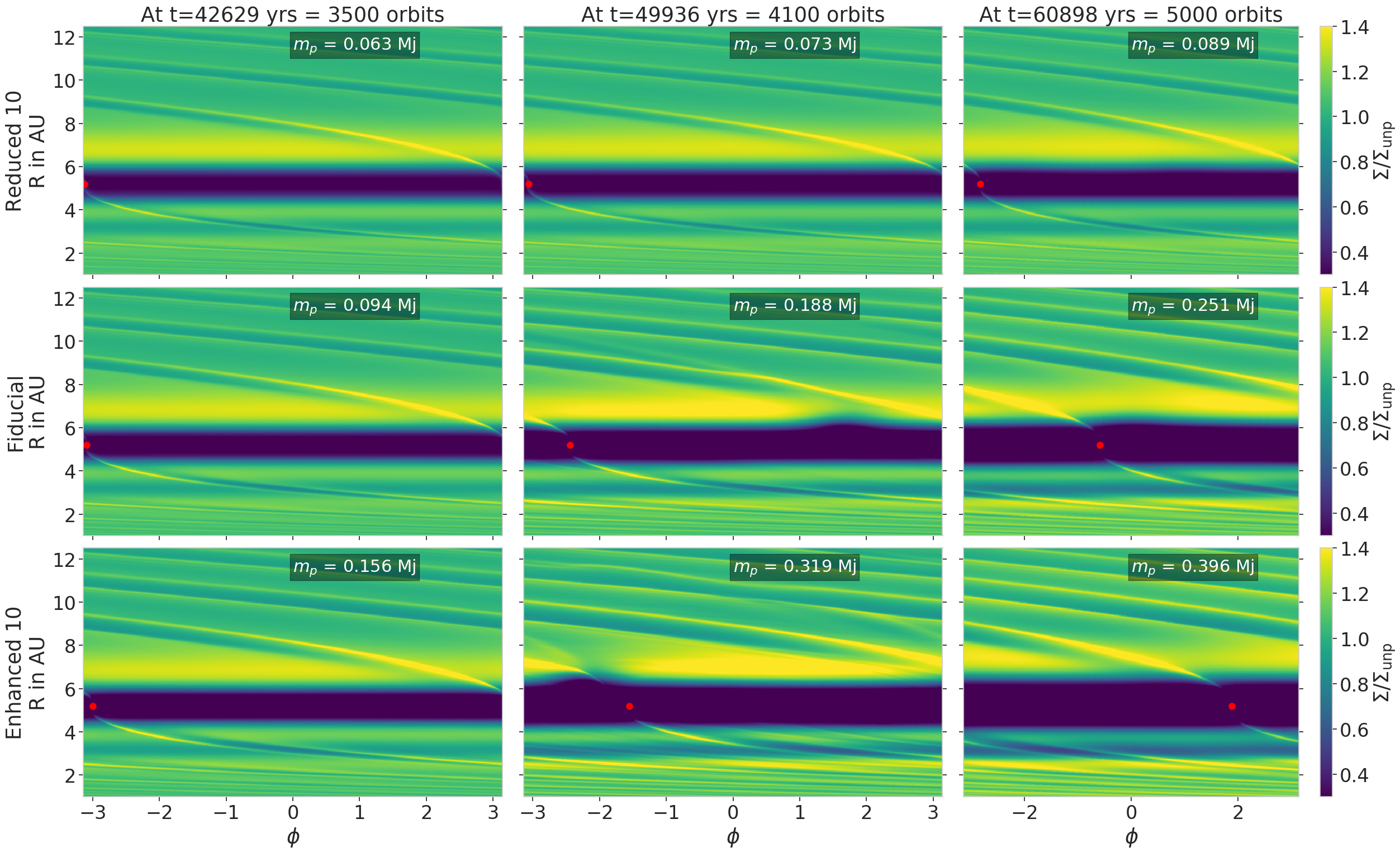}
        \caption{Perturbed surface density at three different time (from left to right) for three planetary accretion rates, with 0.1 (top), 1 (middle) and 10 (bottom) times the nominal accretion rate, in discs with $\alpha = 10^{-4}$, $h = 0.05$. Red dots show the position of the planet. \textit{First row:} density snapshots for the reduced by 10 accretion rate. As the planet is growing really slowly, the gap edges are not very steep, creating a very weak vortex. \textit{Second row:} density snapshots for the fiducial accretion rate. The edges of the created gap are steep enough to trigger the RWI, creating a vortex. The vortex vanishes after $\sim 8.1\times10^{4}$ years. \textit{Third row:} density snapshots for the enhanced by 10 accretion rate. The planet is growing so fast that the edges of the created gap are steep, triggering the formation of a vortex stronger than in the previous cases. The vortex vanishes after $\sim 5.3\times10^{4}$ years.}
        \label{vortices}
\end{figure*}

Another important disc parameter is the viscosity of the disc, which determines how a planet opens a gap \citep{Crida2006,Fung2014,Kanagawa2015} and dictates how gas flows in the vicinity of the planet. Both have a strong impact on gas accretion. In order to determine the influence of viscosity, we run five different simulations with the following alpha parameters: $\alpha = 2\times 10^{-2}$, $10^{-2}$, $10^{-3}$ and $10^{-4}$ (Table \ref{table:1}). As mentioned in Sect. \ref{section_setup}, we need to wait until the initial mass of $\rm 20 \; M_\oplus$ creates a steady gap, defined as a change of less than $1\%$ of the gap depth within 100 planetary orbits. This waiting time is dependent on the disc parameters and is summarized in Table \ref{table:2}.

We can expect the planetary accretion rate to decrease with decreasing viscosity because viscosity dictates how fast gas removed by accretion can be replenished from the viscously spreading disc. This behavior can be seen on the top panels of Fig \ref{differentvisco}: in the first panel, the gas accretion rates are plotted as a function of time. We can see that at lower viscosities, the initial mass needs a longer time to reach a steady gap. This means that at lower viscosity, gas accretion starts later. In the second panel the accretion rates are shifted so that they are plotted as a function of the time after gas accretion is turned on. The accretion rates plotted here show that a lower viscosity results in a lower planetary accretion rate.

At low viscosity, instabilities in the disc are generated \citep{Klahr2003,Fu2014}, resulting in a threshold value for the gas surface density in the vicinity of the planet. If this threshold exists, we would expect a limit of the planetary accretion rate as well. However, in order to resolve these instabilities adequately, high-resolution simulations are needed, which are computationally very expensive. Thus we limit our study to $\alpha \geq 10^{-4}$. 

For $\alpha \lesssim 10^{-4}$, Rossby wave instability is triggered \citep{Lovelace1999,Li2001}. This instability generates vortices at the location of steep density gradients. Such steep gradients are induced by planetary gaps formed at low viscosity \citep{Hammer2017,Pierens2019}, as we can see on the radial density profiles in the lower panel of Fig. \ref{differentvisco}. Vortices will modify the flow of material into the gap, changing the amount of gas available for accretion by the planet. Such impact can be seen in the accretion rates. In the top panel of Fig. \ref{differentvisco}, for $\alpha = 10^{-4}$, the accretion rate is oscillating due to the presence of a vortex at the outer edge of the gap. 
We show in Fig. \ref{vortices}, the 2D ($r, \phi$) surface densities at three different times. The fiducial case is plotted in the middle row. The asymmetrical overdensity (in yellow) located at the outer edge of the gap is characteristic of the presence of the vortex. At later times, the vortex vanishes. In the fiducial case, it completely vanishes after $\rm t \gtrsim 8.1\times10^{4}$ years, inhibiting the oscillations in the accretion rate (top panel of Fig. \ref{differentvisco}).

As the presence of vortices is linked to the steepness of the density gradient, the characteristics of the created vortices depend on how fast the planet grows and creates its gap. \cite{Hammer2017} find that a slowly growing planet will create a weak vortex at the outer edge of the planet gap. We find a similar result when we apply our different accretion rates to simulations with $\alpha = 10^{-4}$. In Fig. \ref{vortices} we show the 2D surface density snapshots for three different accretion rates (reduced by 10, fiducial, and enhanced by 10 -- from top to bottom) at three different times. The overdensity characteristic of the vortex evolves differently whether the planet is accreting quickly or not. We can see that the created vortex at the outer edge of the gap is stronger when the planet features a higher gas accretion rate. As stated previously, we expect to see oscillations in the gas accretion rate, with higher oscillations for the enhanced case as the vortex that is created is stronger.

In the top panel of Fig. \ref{ontostardiffacc_alpha4}, we show the accretion rates reached with the enhanced and reduced accretion rates presented in Sect. \ref{section_diffacc} for $\alpha = 10^{-4}$. As in the fiducial case, all accretion rates show oscillations, with different amplitudes and duration times. As expected, larger accretion rates result in stronger oscillations. The small increase that can be seen on the time evolution of the enhanced accretion rates from $t \simeq 4.1\times10^{4}$ years to $\sim 5.1\times10^{4}$ years, is also due to the presence of vortices. It corresponds to the moment when the vortices are the strongest. Once they vanish after $\sim 5.3\times10^{4}$ years, the oscillations in the accretion rates vanish too. Our results are, therefore, in agreement with \cite{Hammer2017}, with the vortices pushing material into the vicinity of the planet and their presence demonstrating a link to an enhancement in the gas accretion rate.

The presence of these vortices has a strong impact on the density profiles of the disc. The large viscous timescale at low viscosity results in a long period of time for the disc to adjust to the presence of the planet, resulting in an highly perturbed disc structure. The perturbed density profiles are shown in the lower panel of Fig. \ref{ontostardiffacc_alpha4}. Perturbations are particularly high in the inner disc. This has a strong influence on the stellar accretion, shown on the middle panel of Fig. \ref{ontostardiffacc_alpha4}. Compared to what we describe in Sect. \ref{section_ontostar}, the scaling of the stellar accretion with the planetary accretion is reversed here: a more efficiently accreting planet results in a larger stellar accretion rate. We expect this effect to flip at later time, when the inner disc gets depleted by the viscosity: the viscous time at $\alpha = 10^{-4}$ is 100 times longer than for $\alpha = 10^{-2}$, requiring $\rm t_\nu = 8 \; Myrs$ for material to move from the position of the planet to the inner disc, which is too long for computational integration. Therefore, the depletion of the inner disc is less affected by smaller viscosities: instead the inner disc is mostly influenced by the presence of the growing planet. Larger planets induce larger perturbations: the enhancement \footnote{We note that we plot the perturbed surface density, therefore the "bump" shown here is not present in the actual gas surface density profile. This enhancement in density is caused by the growing planet that pushes material from its orbit.} created by the planet in the inner disc, at $r < 1$ AU as shown in the lower panel of Fig. \ref{ontostardiffacc_alpha4}, is larger for the largest planet, that is, in the case of enhancement by 10. The presence of this enhancement at the inner disc is the reason the stellar accretion is higher compared to the disc without a planet. As soon as the disc starts to deplete its inner disc via viscous spreading, the accretion rate onto the star behaves as displayed in Fig. \ref{ontostardiffacc}, where we observe a higher decrease in stellar accretion rate with quickly accreting planets. This behavior was confirmed by investigating the evolution of the stellar accretion rate for an intermediate viscosity: for $\alpha = 10^{-3}$ (not shown), the stellar accretion rate is larger compared to a disc without a planet at the beginning of the simulation, but it then becomes smaller when the inner disc starts to deplete via viscous spreading.

Viscosity has an impact on how the planet will open a gap, as can be seen on the lower panel of Fig. \ref{differentvisco}: depending on the viscosity, a gap is opened at a different mass and the gap has a different shape. In the next section, we investigate the impact of gas accretion on the gap-opening mass.

\subsection{Gap-opening mass} \label{section_gapopeningmass}

\begin{figure}[t]
   \centering   
   \includegraphics[scale=0.255]{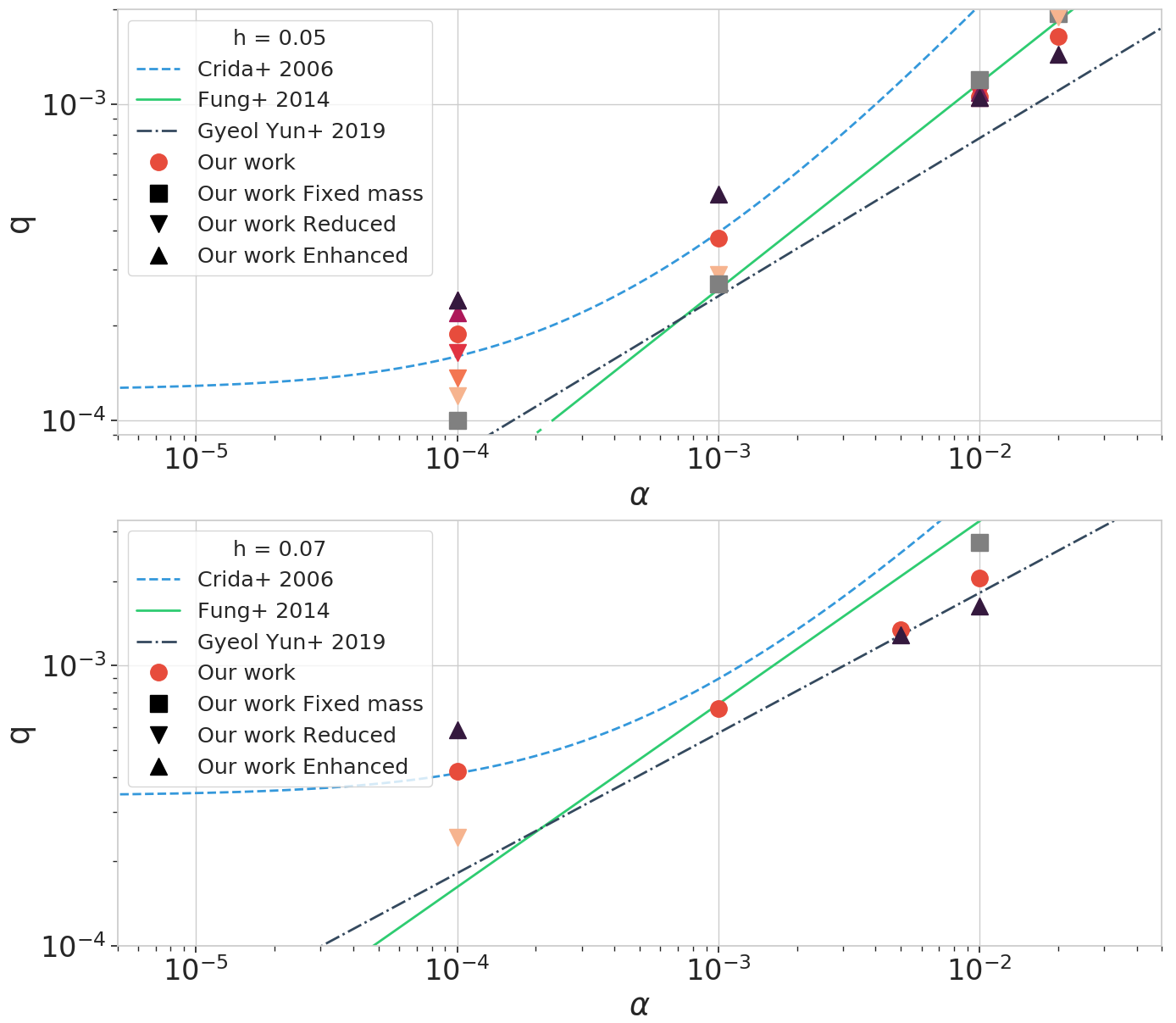}
   \caption{Gap-opening mass as a function of the viscosity for different criteria and our simulations. The lines represent the different gap-opening criteria from the literature: \cite{Fung2014} in solid green, \cite{Crida2006} in dashed blue, and \cite{Yun2019} in dashed-dotted black. The dotted green line represents the mass limit of the \cite{Fung2014} study. None of these works include planetary gas accretion. Red dots represent our fiducial simulations. Results from the different accretion rates are represented by upward triangles for the enhanced accretion rates and downward triangles for the reduced accretion rates. The color coding is the same as in Fig. \ref{diffacc}: darker colors represent larger accretion rates. The two different panels are for the two high aspect ratios: $h$ = 0.05 (top) and 0.07 (bottom). The initial mass of the planet is $20 M_\oplus$. Gas accretion has different impact as a function of disc parameters. This discrepancy can be explained by the time needed to the disc to react to a change of planetary mass.}
   \label{gapopeningmass}
\end{figure}

\begin{figure}[t]
   \centering   
   \includegraphics[scale=0.255]{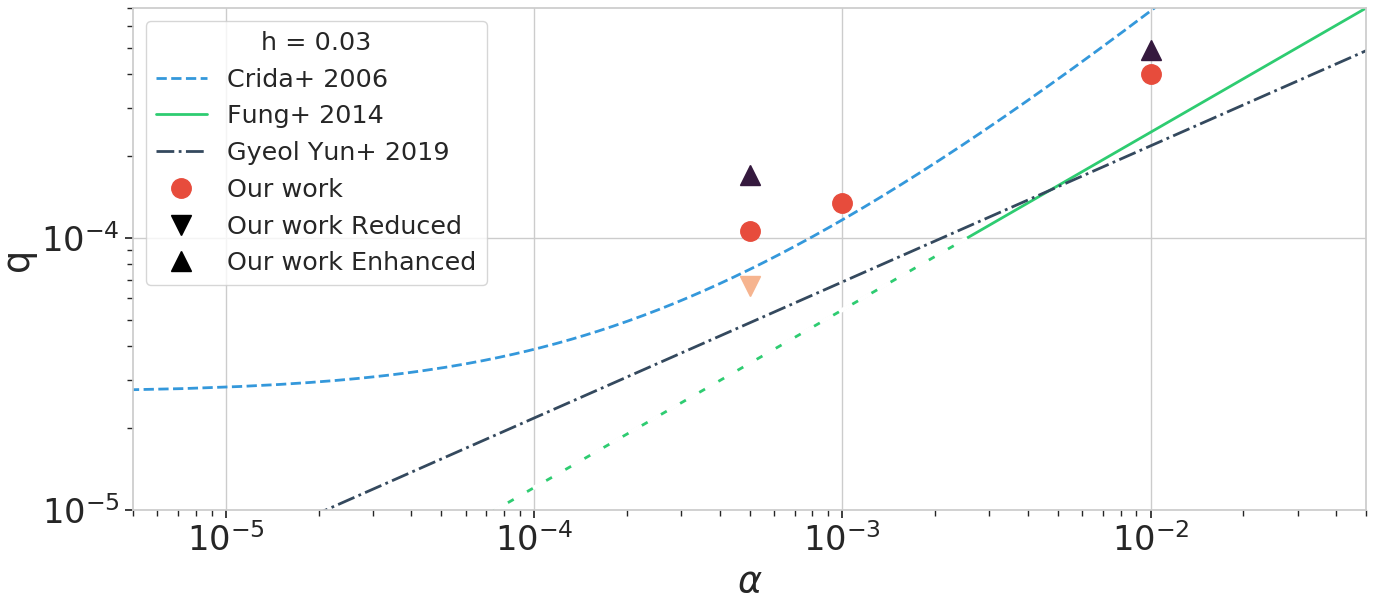}
   \caption{Same as Fig. \ref{gapopeningmass} for h = 0.03. The initial mass of the planet is of 10 $M_\oplus$, changing the gas accretion rate and, therefore, the gap-opening mass compared to the other simulations where $m_{\rm init} = 20 M_\oplus$.}
   \label{gapopeningmassh3}
\end{figure}

The criteria for gap-opening mass have been derived in several previous studies. In \cite{Crida2006}, they defined the gap-opening mass as the mass at which $\Sigma_{\rm gap} = 0.1 \; \Sigma_{\rm unp}$, which we also use in our work. The derived gap-opening criterion depends on the aspect ratio and viscosity:

\begin{equation}
    P = \frac{h}{q^{1/3}}+50\frac{\alpha h^2}{q} \leqslant 1   
    \label{eq3}
,\end{equation}

\noindent where q is the planet to star mass ratio and a planet opens a gap when $\rm P \leqslant 1 $. This criterion has been tested for viscosities  of $10^{-5} \leqslant \alpha \leqslant 3\times10^{-1}$ and aspect ratios of $0 \leqslant h \leqslant 0.3$ by \cite{Crida2006}. Here we focus on the gap-opening mass, that is, the mass at which q we satisfy $\rm P = 1$ as a function of the viscosity and aspect ratio. We plot this formula in the q-$\alpha$ space in Figs. \ref{gapopeningmass} and \ref{gapopeningmassh3} for the three different aspect ratios studied in this work. 

In other studies, equations for $\Sigma_{\rm gap}$ were derived. In \cite{Kanagawa2015}, the depth of the gap is given as:
\begin{equation}
    \frac{\Sigma_{\rm gap}}{\Sigma_0} = \frac{1}{1 + 0.04K}
    \label{eq4}
,\end{equation}

\noindent where $K = q^2h^{-5}\alpha^{-1}$. This criterion is valid for $K \leqslant 10^{4}$. For the mass range given on each axis, this gives a validity range from $\alpha = 2\times10^{-6}$ to $\alpha = 1.2\times10^{-3}$ for $h = 0.05$ and from $\alpha = 6\times10^{-7}$ to $\alpha = 5\times10^{-4}$ for $h = 0.07$, which is, therefore, valid for the low viscosities given here. This was derived for a uniform disk, meaning that the initial surface density was constant in radial direction. \cite{Yun2019} added a correction to this work by studying gap opening in a non-uniform disk, having a power law initial surface density profile with an exponential cutoff. They found a very similar result as that found by \cite{Kanagawa2015}, with $\Sigma_{\rm gap}/\Sigma_0 = 1/(\rm 1 + 0.046K)$. To be able to compare this corrected criterion with Crida's, we looked for the mass at which $\Sigma_{\rm gap }/\Sigma_0 = 0.1$. We plot this as a function of the viscosity for different aspect ratios in Fig. \ref{gapopeningmass}. 

\cite{Fung2014} also derived a gap-opening criterion for planetary masses in the range of $10^{-4} \leqslant q \leqslant 5\times10^{-3}$, for $10^{-3} \leqslant \alpha \leqslant 10^{-1}$ and for $0.04 \leqslant h \leqslant 0.1$:
\begin{equation}
    \frac{\Sigma_{\rm gap}}{\Sigma_0} = 0.14 \Big(\frac{q}{10^{-3}}\Big)^{-2.16} \Big(\frac{\alpha}{10^{-2}}\Big)^{1.41} \Big(\frac{h}{0.05}\Big)^{6.61}
    \label{eq5}
.\end{equation}

In Fig. \ref{gapopeningmass}, we see that Fung and Kanagawa's criteria show a linear evolution for the gap-opening mass as a function of the viscosity, unlike Crida's criterion that levels off at low viscosity. This leveling off is caused by the fact that even when the viscosity vanishes, the planet is still in competition with the pressure of the disc to create a gap, creating a lower threshold for the gap-opening mass. Moreover, there should be a threshold at low viscosity as the presence of vortices generates a certain background level of turbulence, independently of the prescribed alpha viscosity. 

In order to determine the influence of planetary gas accretion on the gap-opening mass within our simulations, we investigated the gap-opening masses of fixed mass planets in our simulations (squares in Fig. \ref{gapopeningmass}). Our planets with fixed mass seem to be opening gaps following Fung's criterion for $h = 0.05$ and high viscosity only. For the other aspect ratios, all gap-opening criteria seem to be failing. We suspect that our simulations do not match  any criteria directly as they are all derived from fits to simulations and, a a result, inducing errors. Using the gap-opening masses derived from the \cite{Crida2006} criterion (Eq.\ref{eq3}) we actually find deeper gaps than  $0.1 \; \Sigma_{\rm unp}$. It is also important to note that we are investigating gap-opening mass in non-equilibrium discs: the mass accretion rate onto the star is evolving over time as material is accreted onto the planet and the star. Earlier studies \citep{Crida2006,Fung2014,Kanagawa2015} were made on the basis of equilibrium discs, meaning that the stellar accretion rate is constant over time. Having a viscously evolving disc may change the gap-opening mass because the accretion onto the star leads to depletion in the inner disc, influencing the material around the planet and therefore influencing how the gap is opened (especially at high viscosities). To maintain consistency, we therefore compare gap-opening masses within our simulations.

\begin{figure}[t]
   \centering   
   \includegraphics[scale=0.26]{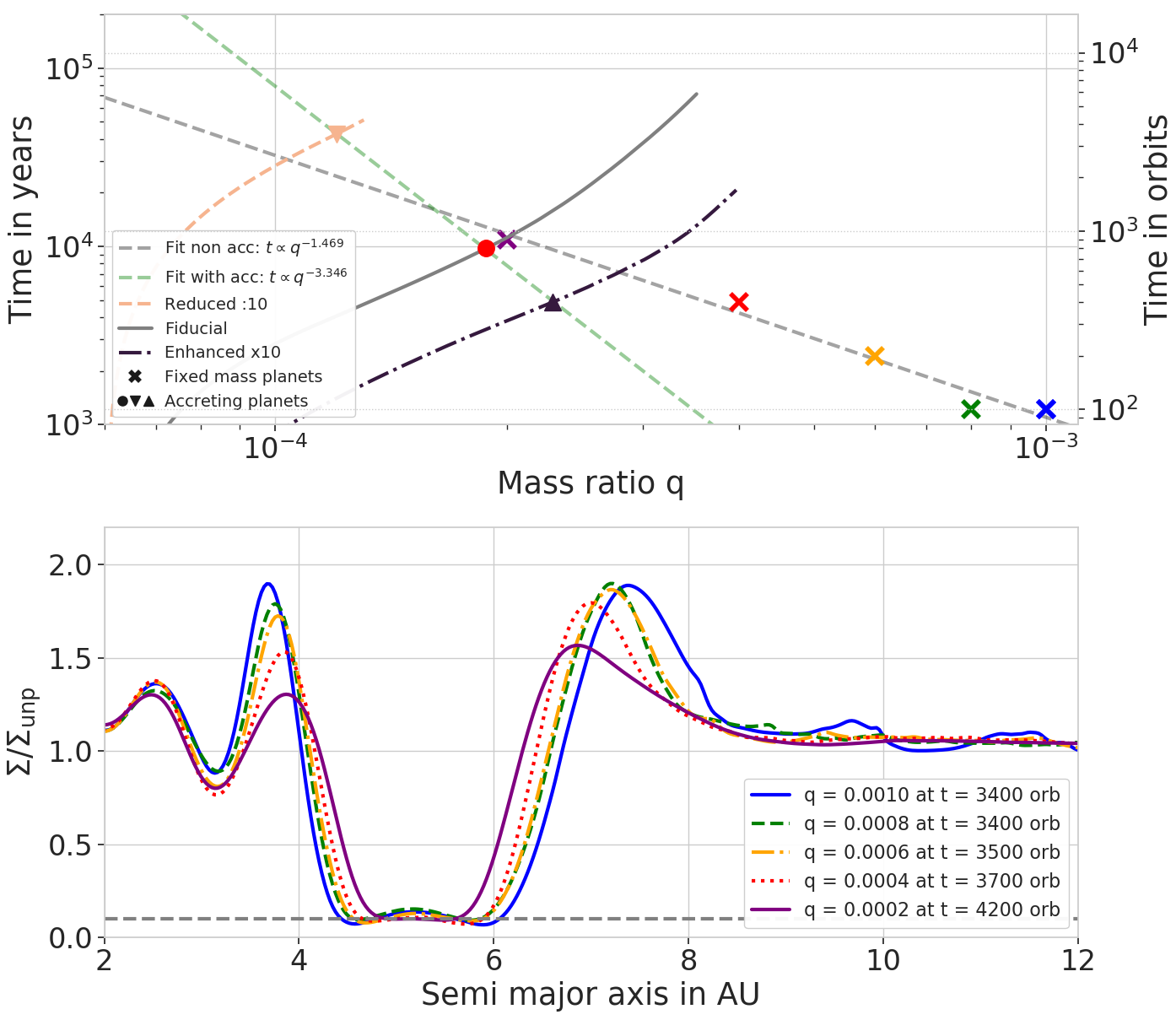}
   \caption{Time needed for gap opening as a function of the planetary mass for $\alpha = 10^{-4}$ and $h=0.05$. The gap opening time is defined as the time needed to reach the $\Sigma/\Sigma_{\rm unp} = 0.1$ threshold by gas accretion, therefore, we removed the time needed to reach the initial gap, $t_{\rm gap,init}$ (see Table \ref{table:2}). \textit{Top:} Gap opening time as a function of the planetary mass. Fixed-mass planets are represented by crosses and linearly fitted by the dashed gray line. For comparison, the time evolution of the planetary masses for the accretion rate enhanced by 10, fiducial accretion rate, and that reduced by 10 are represented by the dashed-dotted dark, solid gray, and dashed light lines, respectively. Their gap-opening masses are represented with the same symbol as in Fig. \ref{gapopeningmass}. For these disc parameters, the gap-opening mass and time of the accreting planet follow the predictions of the fixed-mass planets. The disc response time is therefore the dominant phenomenon for gap opening. \textit{Bottom:} Perturbed surface densities of the fixed-mass planets at gap opening time. Gap-opening mass and time are defined by the moment at which the surface density is $\Sigma = 0.1\Sigma_{\rm unp}$.}
   \label{time_gap_open_a4}
\end{figure}

\begin{figure}[t]
   \centering   
   \includegraphics[scale=0.26]{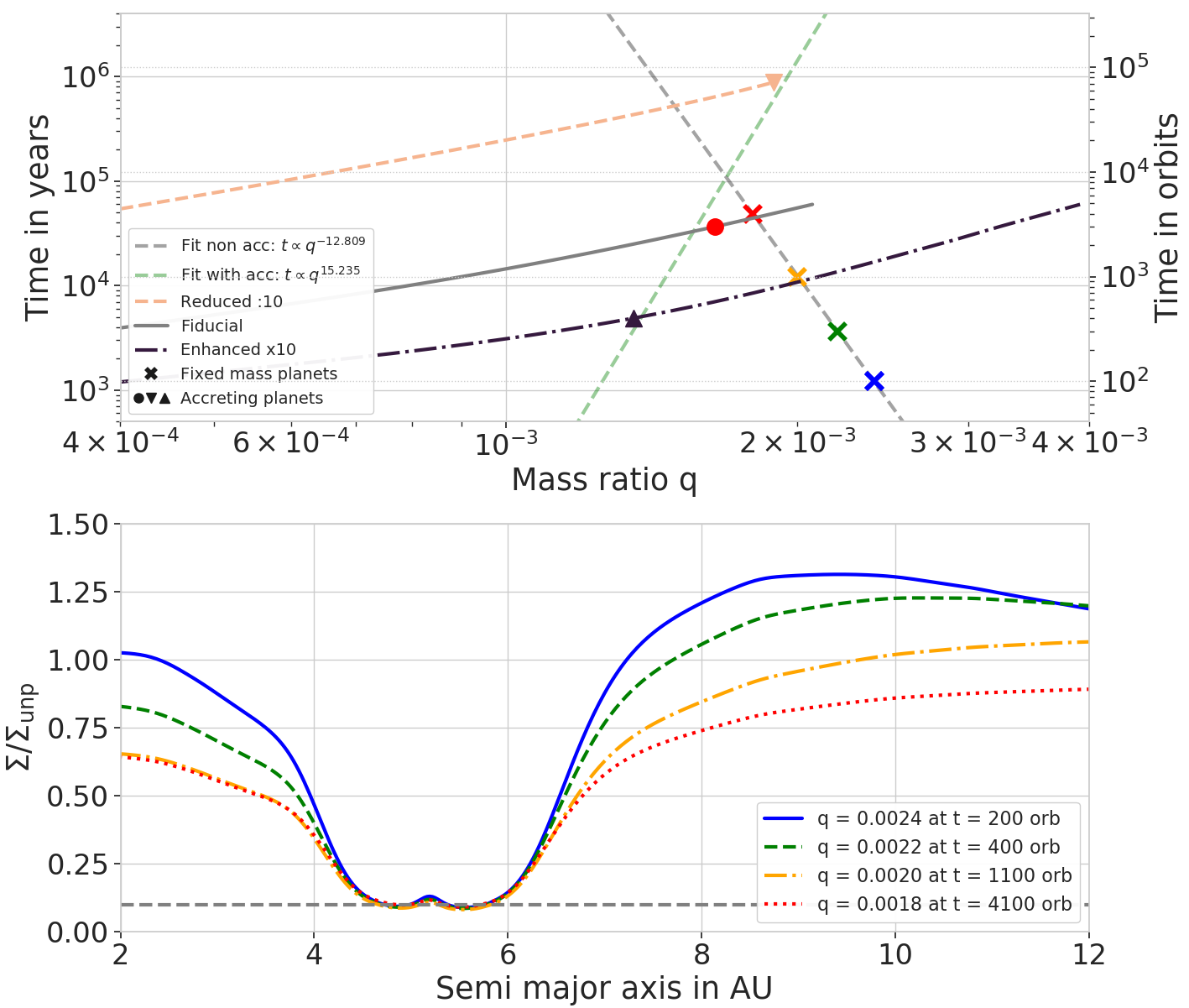}
   \caption{Time needed for gap opening as a function of the planetary mass for $\alpha = 2\times10^{-2}$ and $h=0.05$. Panels represent the same quantities as in Fig. \ref{time_gap_open_a4}. For these disc parameters, the gap-opening mass and time of the accreting planets do not follow predictions for the fixed-mass planets. Gas accretion is, therefore, the dominant phenomenon for gap formation.}
   \label{time_gap_open_a2}
\end{figure}

At this point in the study, we can identify three behaviors depending on the level of viscosity: 1) at low viscosity (e.g., at $\alpha = 10^{-4}$ for $h=0.05$), the gap-opening mass is highly dependent on the gas accretion rate. Simulations with faster accreting planets result in larger gap-opening masses, while simulations with slower accreting planets result in smaller gap-opening masses; 2) at high viscosity (for example at $\alpha = 2\times10^{-2}$ for $h=0.05$), we find the opposite behavior. A simulation with a high accretion rate will open a gap at a smaller mass than a slowly accreting planet; 3) as the behavior flips between high and low viscosity, a peculiar viscosity exists, for example at $\alpha = 10^{-2}$  for $h=0.05$, for which the gap-opening mass is insensitive to the different gas accretion rates. For the seven simulations, the gap opening mass is nearly the same: $\rm m_{gap} \simeq 1.06 \; \rm M_J$.

In order to explain these behaviors, we investigate the time needed for a fixed mass planet to open a gap as a function of its mass, at low and high viscosities. Our results are shown in Figs. \ref{time_gap_open_a4} and  \ref{time_gap_open_a2} for low ($\alpha = 10^{-4}$) and high ($\alpha = 2\times10^{-2}$) viscosities, respectively. The crosses represent the time needed for fixed-mass planets of different masses to open a gap. To be consistent with regard to the comparison, in this case, the fixed-mass planets are introduced in exactly the same way as the accreting planets: after the $20M_\oplus$ core is introduced and its initial gap has been created, the mass of the planet is switched to the final mass of interest. The time needed to open a gap corresponds then to the time needed to reach the $\Sigma/\Sigma_{\rm unp}= 0.1$ threshold after the planet mass is switched to its final mass.

As expected, we see that more massive planets open gaps more rapidly than low-mass planets \citep{LinPapaloizou1986}. In consequence, the way the planet is introduced in the disc has a strong influence on the gap-opening mass (more precisely, on the time at which the $0.1\;\Sigma_{\rm unp}$ threshold is reached). To make a comparison with accreting planets, we show, using the same figures, the time evolution of the masses of the accreting planets. The dots and triangles represent the time and mass at which each accreting planet is observed to open a gap (same dots and triangles of Fig. \ref{gapopeningmass}). On the lower panels of Figs. \ref{time_gap_open_a4} and \ref{time_gap_open_a2}, we show the surface density profiles of the non-accreting planets at the time and mass used for the gap definition. We observe two different behaviors at low and high viscosities that can explain the gap-opening masses of our accreting planets (Fig. \ref{gapopeningmass}). First, at low viscosity (Fig. \ref{time_gap_open_a4}), the gap-opening masses of the accreting planets and the one of the fixed-mass planets are both inversely proportional to the planet mass ($t \propto q^{-X}$). In this case, the gap-opening phenomenon is governed by the disc response to the change in the mass of the planet. Gas accretion only changes the value of the slope compared to the fixed-mass planets ($X\simeq-1.47$ for the fixed-mass planets and $X\simeq-3.35$ for the accreting planets). This means that the process of gas accretion is dominated by the disc response time, that is, the time needed for the disc to react to a change of planetary mass. In this case, when the planet is slowly introduced in the disc, it opens a gap at a lower mass than a planet that is rapidly introduced. This explains the behavior seen in Figs. \ref{time_gap_open_a4} and \ref{gapopeningmass} at low viscosity: slowly accreting planets are slowly changing in mass, resulting in smaller gap-opening masses.
On the other hand, at high viscosity (Fig. \ref{time_gap_open_a2}), the behavior is completely different. The gap-opening masses of the fixed-mass planets is still inversely proportional to the planet mass ($t \propto q^{-X}$) but the gap-opening mass of the accreting planets is proportional to $q$ ($t \propto q^{+X}$). In this case, the disc response is quick enough for planetary gas accretion to help carve a deeper gap. Gap opening is, therefore, dominated by gas accretion and not by the disc response time. 

The key parameter here is the disc response time to a change of planetary mass. A formula for the gap opening time in an inviscid disc was derived by \cite{LinPapaloizou1986}. They found that $\tau_{\rm gap} \propto q^{-2}$. As plotted in Figs. \ref{time_gap_open_a4} and  \ref{time_gap_open_a2}, we find $\tau_{\rm gap} \propto q^{-1.47}$ for $\alpha = 10^{-4}$ and $\tau_{\rm gap} \propto q^{-12.81}$ for $\alpha = 2\times10^{-2}$, meaning that at higher viscosity, the gap opening time is more dependent on the planetary mass. It is important to mention that the discrepancy between our result and that of \cite{LinPapaloizou1986} comes from the importance of viscosity in the gap-opening process and the definition for the gap opening time. Indeed, we define the gap-opening mass as the mass needed to create a gap of a certain depth ($\Sigma/\Sigma_{\rm unp} = 0.1$), whereas they define it on the basis of when the forces applied by the planet on the disc are equilibrated and vice versa.

In conclusion, gas accretion has a different impact on the gap-opening mass depending on the disc response time. It is a dominating phenomenon at high viscosities. These differences in gap-opening mass can have an important impact on the transition to type II migration, which is investigated in the next section.

\begin{figure*}[t]
   \centering   
   \includegraphics[scale=0.30]{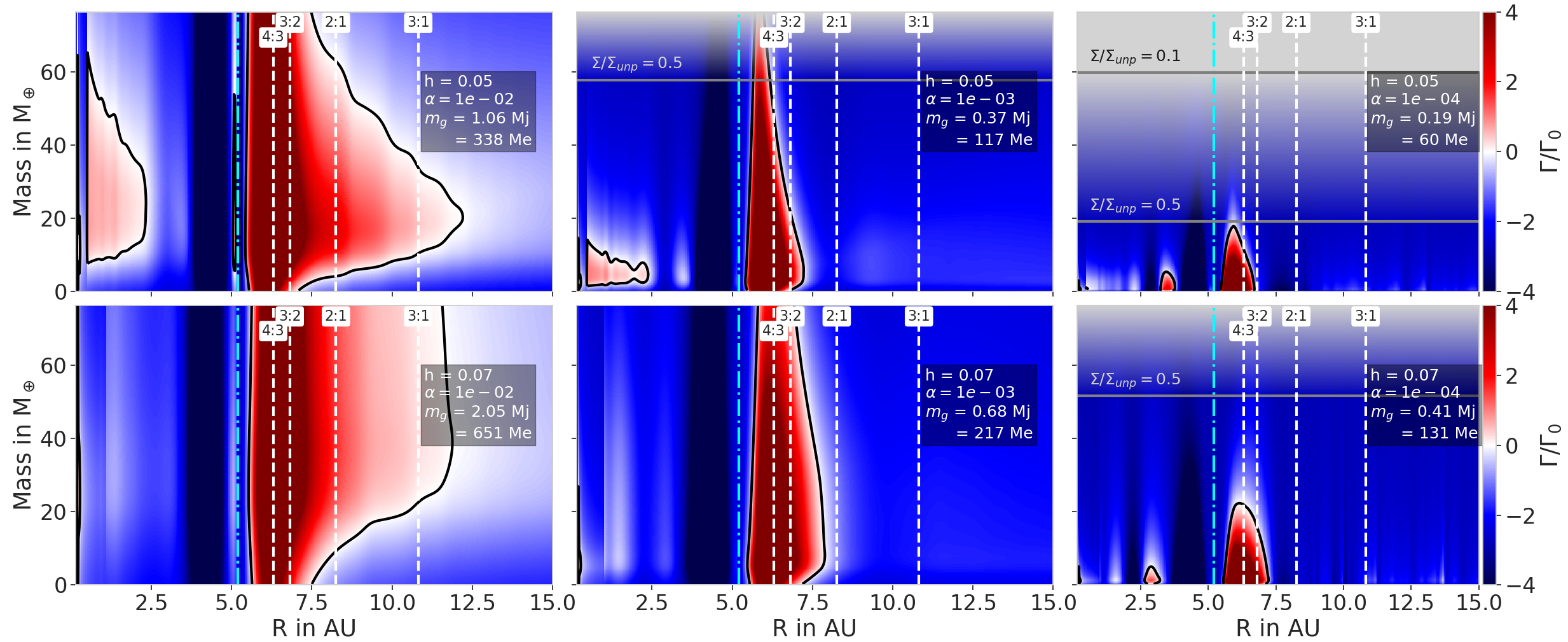}
   \caption{Migration maps for low-mass planets orbiting near our accreting planets. The accreting planet has reached gap-opening mass. The different panels represent different disc parameters: the viscosity decreases from left to right ($\alpha = 10^{-2}, 10^{-3}, 10^{-4}$) and the aspect ratio increases from top to bottom ($h = 0.05,0.07$). Positive (red) torques indicate outward migration while negative (blue) torques represent inward migration. The black solid line represents the zero torque line, a position where the second planet would stop migrating. The two gray horizontal lines represent the masses at which $\Sigma/\Sigma_{\rm unp} = 0.5$ and $0.1$. We consider that the planet will smoothly switch from type I to type II migration between these two masses. The vertical white dashed lines locate the positions of the resonances with the accreting planet situated at 5.2 AU (vertical blue dotted dashed line).}
   \label{migrationmaps}
\end{figure*}

\begin{figure*}[t]
   \centering   
   \includegraphics[scale=0.30]{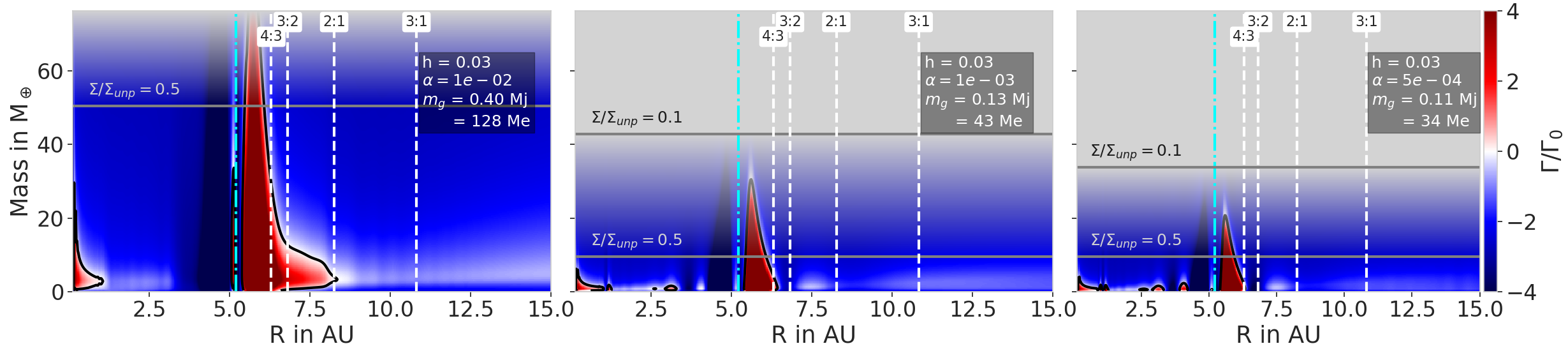}
   \caption{Same as Fig. \ref{migrationmaps} but for $h=0.03$. Here, $\rm m_{init} = 10 \;M_\oplus$ and $\alpha = 10^{-2},10^{-3},$ and $5\times10^{-4}$.}
   \label{migrationmapsh3}
\end{figure*}

\subsection{Migration maps} \label{section_migrationmap}

Planets embedded in discs interact with the gas by exchanging angular momentum, which results in planet migration (for recent reviews see \citealt{Kley2012,Baruteau2014}). Migration is a robust phenomenon and a central ingredient in global models of planet formation that attempt to reproduce known exoplanet systems as well as our own Solar System (for a review, see \citealt{Raymond2018}). In order to study the type I migration of planets that could be present in the same disc as our gas-accreting planets, we employ the formulae from \cite{Paardekooper2011} to estimate the torque acting on the smaller bodies. In their work, they derived a formula based on the temperature gradient $\rm T \propto r^{-\beta_T}$ and surface density gradient $\rm \Sigma \propto r^{-\alpha_\Sigma}$. The total torque is composed of five components coming respectively from the Lindblad torque, the barotropic and entropic horseshoe drag torques and the barotropic and entropic linear corotation torque:
\begin{equation}
\begin{split}
        & \rm \gamma \Gamma_{L}/\Gamma_0 = \rm  -2.5-1.7\beta_T+0.1\alpha_\Sigma , \\
        & \rm \gamma \Gamma_{HS,baro}/\Gamma_0 = \rm 1.1(3/2 - \alpha_\Sigma), \\ 
        & \rm \gamma \Gamma_{lin,baro}/\Gamma_0 = \rm 0.7(3/2 - \alpha_\Sigma), \\
        & \rm \gamma \Gamma_{HS,ent}/\Gamma_0 = \rm 7.9 \xi /\gamma, \\ 
        & \rm \gamma \Gamma_{lin,ent}/\Gamma_0 = \rm (2.2 - 1.4/\gamma)\xi 
\end{split}
,\end{equation}
where $\gamma = 1.4$ is the adiabatic index and $\rm \Gamma_0 = (q/h)^2\Sigma_pr_p^4\Omega_p^2$ is the normalizing torque depending on $\rm q$ (the mass ratio between the planet and the star), $h$ is the aspect ratio, $\rm \Sigma_p$ is the surface density at the location of the planet, $\rm r_p$ is the distance star-planet, and $\rm \Omega_p$ is the planet's angular velocity. The total torque $\rm \Gamma_{tot}$ is a linear combination of these torques where the coefficients $F$, $G,$ and $K$ depend on the saturation parameters $p_\nu$ and $p_\chi$:
\begin{equation}
\begin{split}
    \Gamma_{\rm tot} & = \Gamma_{\rm L} + F(p_\nu)G(p_\nu)\Gamma_{\rm HS,baro} + (1-K(p_\nu))\Gamma_{\rm lin,baro} \\ 
    & + F(p_\nu)F(p_\chi)\sqrt{G(p_\nu)G(p_\chi)}\Gamma_{\rm HS,ent} \\
    & + \sqrt{(1-K(p_\nu))(1-K(p_\chi))}\Gamma_{\rm lin,ent}
\end{split}
,\end{equation}

\noindent where $p_\nu = 2/3\sqrt{(r_p^2\Omega_px_s^3)/(2\pi\nu_p)}$ is the viscous saturation parameter depending on the horseshoe half width $x_s$ as described in \cite{Masset2006} and the viscosity, $\nu_p$, at the location of the planet and $p_\chi = 3p_\nu/2 \sqrt{\nu_p/\chi_p}$ is the thermal saturation parameter depending on the thermal diffusion, $\chi_p$, which depends on the opacities in the disc, where we use the \citet{Bell&Lin1994} opacities.
This torque formula from \cite{Paardekooper2011} has been compared to 3D radiation hydrodynamic simulations \citep{Bitsch&Kley2011,Lega2015} and found to be a good match.

A negative torque would result in an inward migration, towards the star. On the other hand, a positive torque reflects outward migration, toward the outer parts of the disc. Inward migration is mostly due to the Lindblad component of the torque. In order to have outward migration in our locally isothermal case, where the temperature gradients are quite shallow, sharp positive radial gradients in the surface density are needed \cite{Masset2006}. These gradients appear at the outer edge of the gap created by the accreting planet. 

In Figs. \ref{migrationmaps} and \ref{migrationmapsh3}, we show the migration maps for different viscosities and different aspect ratios when the accreting planet has reached its gap-opening mass. Regions of inward (blue) and outward (red) migration are represented as well as the potential equilibrium position of the low mass migrating planet: the zero-torque position (black line) and the resonances (dashed white vertical lines) with the accreting giant planet, whose position is marked by the dashed blue line.

An outer lower mass planet would stop migrating either at the zero-torque location or if it is trapped in resonance with the inner planet, depending on its migration speed \citep{Thommes2005,Pierens&Nelson2008}. In our simulations, if an inward migrating planet can jump the 2:1 resonance and continue to migrate further in, it would reach the position of the 3:2 resonance for the lowest viscosities, when the torques saturate. At high viscosity, for all aspect ratios, the outward migration zone is wide enough to overlap over the 3:2 resonances positions and the 2:1 resonances positions if the mass of the second planet is small enough. It these cases, the migrating planet is prevented to be locked in 3:2 resonance by the zero torque line, specifically for low-mass planets. When the planet mass is small enough, the 2:1 resonance is also unreachable due to the extend of the outward migration zone. Depending on the shape of the gap created and the migration speed of the low-mass planet, the capture in certain resonances might therefore be avoided. The capture in resonance is important for the grand tack scenario  \citep{Walsh2011}, where Jupiter and Saturn get locked in 3:2 or 2:1 resonance and start migrating outward \citep{Masset&Snellgrove2001,Morbidelli&Crida2007,Pierens&Nelson2008,Raymond2011,Pierens2014,Chametla2020}. We discuss, in Sect. \ref{section_discussion_structure}, the impact of resonance capture on the structure of our Solar System.

\cite{Pierens2014} investigate which disc parameters ($\Sigma_0$ and $\alpha$) and capture in resonance are needed to allow outward migration of Jupiter and Saturn. For h=0.05 (middle row in Fig. \ref{migrationmaps}) and our surface density ($\Sigma/\Sigma_{\rm MMSN} = 1.5$)\footnote{Note that their surface density profile is different from our setup: $\Sigma \propto r^{-3/2}$ as opposed to $\Sigma \propto r^{-1}$ in this work. Therefore the outcome of planet migration in our case might be different from their results.} with $\alpha=10^{-3}$, outward migration for the Jupiter-Saturn pair would happen in the 3:2 resonance. Still, according to \cite{Pierens2014}, the Jupiter-Saturn system would show divergent migration for $\alpha=10^{-2}$ and be captured in the stable 2:1 configuration for $\alpha=10^{-4}$. In Fig. \ref{migrationmaps}, we see that the capture in the 3:2 resonance is possible for $\alpha = 10^{-3}$ and $h = 0.05;$  for these disc parameters, Saturn would not open a deep gap ($\rm m_S = 95 M_\oplus < m_{\rm gap} = 117 \rm M_\oplus $). Therefore, our migration maps agree with the results found in \citep{Pierens2014}, assuming that Saturn's migration speed allows it to be locked in the 3:2 resonance. In addition, our results show that N-body simulations of growing planets should take the disc profile into account to correctly assess the migration behavior of the planets.

\begin{figure*}[t]
   \centering   
   \includegraphics[scale=0.29]{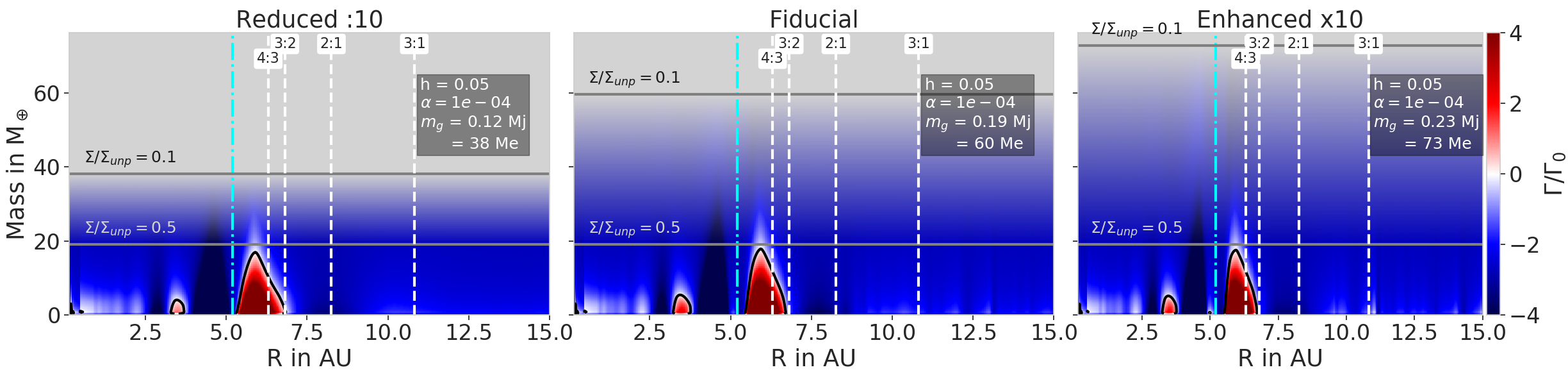}
   \caption{Migration maps for $\alpha = 10^{-4}$ and $h = 0.05$ for low-mass planets orbiting near our accreting planets. The accreting planet has reached the gap-opening mass. The three panels represent different accretion rates: the reduced by 10 (left), fiducial (middle), and enhanced by 10 (right). The plotted information is the same as in Fig. \ref{migrationmaps}. As gas accretion has an influence on the gap-opening mass, the migration maps look different due to the switch from type I to type II migration when a gap is opened. Here, the $\Sigma/\Sigma_{\rm unp} = 0.5$ is at the same mass for all the different accretion rates as the initial depth for these discs parameters is smaller than 0.5 (see Table \ref{table:2}).}
   \label{migrationmaps_a4}
\end{figure*}

As a planet grows and carves a deep gap, it is transitioning from type I to type II migration. In Fig. \ref{migrationmaps_a4}, we show the migration maps for $h = 0.05$ and $\alpha = 10^{-4}$ for the fiducial, the enhanced, and the reduced by 10 rates for planetary gas accretion. If the planet accretes slowly, a gap is opened at a smaller mass than when the planet accretes quickly for these disc parameters (see Sect. \ref{section_gapopeningmass}). The mass at which the planet is expected switch from type I to type II migration (depicted by the gray area in Figs. \ref{migrationmaps}, \ref{migrationmapsh3}, and \ref{migrationmaps_a4}) is, therefore, dependent on the planetary accretion rate, assuming that the second potential planet in the system accretes at the same rate and would thus have the same gap-opening mass. At these disc parameters, the difference in gap-opening mass does not have an impact on the capture in resonance: the region of outward migration is small and does not overlap with resonances for planets more massive than $15 \; M_\oplus$.
The dynamics of multiple planetary systems can be highly impacted by planetary gas accretion via the influence on the migration type and on the potential trapping in resonance.

\section{Discussion}
\label{section_discussion}

\subsection{Accretion onto the star} \label{section_discussion_accontostar}

In this work, we study\ the influence of planetary gas accretion on the stellar accretion (Sects. \ref{section_ontostar} and \ref{section_diffvisco}). We show that the stellar accretion is reduced with increasing planetary accretion rates. Even though, at low viscosity, our results show an enhancement of the stellar gas accretion rate, we expect this trend to flip and follow the high viscosity case after reaching the viscous time needed for material to reach the inner disc from the planet position.

In both previous cases, the stellar accretion was only decreased or increased by a factor of up to three compared to the case of a disc without any planet. These results are quite different from what \cite{Manara2019} find. In their models, they find that stellar gas accretion rate can be reduced by over two orders of magnitude due to the presence of giant planets. In our simulations, these large spreads of stellar accretions could only be achieved by changing the disc viscosity over orders of magnitude, however, this parameter is fixed in the simulations used in \cite{Manara2019}. The difference might come from the difference in gas flow between a 2D and a 1D disc: indeed, in the approach of \cite{Manara2019}, gas accretion onto the star is derived from the viscous spreading of a 1D disc containing a giant planet. As \cite{Lubow2006} conclude in their work, in a  2D disc, mass can flow through the gap even in the presence of a giant planet. Moreover, in \cite{Manara2019}, the planetary gas accretion rates are higher than ours as theirs are proportional to the unperturbed surface density in their accretion routine \citep{Mordasini2012}. Therefore, the 1D model from \cite{Manara2019} might be overestimating the efficiency of the planet in blocking material from the inner disc.

In addition, \cite{Lubow2006} showed that the modeling of the inner disc and the process whereby gas is accreted onto the planet is of crucial importance for calculating the stellar accretion rate. Our simulations indicate that planetary gas accretion might have a smaller impact than expected on the stellar gas accretion rates.

\subsection{Implications for observations} \label{section_discussion_observations}

Observations of protoplanetary discs have revealed structures in the dust distribution \citep{Pinilla2012,Andrews2018,Zhang2018}. However, these dust structures are ultimately caused by the drifting dust grains trapped at pressure bumps in the gas discs. Therefore, as mentioned in previous sections, looking at the pressure bumps created by planets in the gas can give us an insight into how the dust is trapped and how gas accreting planets might influence the interpretation of the observations. 
When we compared the pressure bumps in Sects. \ref{section_accvsnonacc} and \ref{section_diffacc} for our simulations with the fiducial parameters, we found that the pressure bumps were too similar to show distinguishable features. This result is expected, as in Sect. \ref{section_gapopeningmass} we showed that for these disc parameters, gas accretion has almost no influence on the gap depth. On the other hand, at low viscosity the gap-opening mass changed dramatically with the planetary accretion rate.

\begin{figure}[t]
   \centering   
   \includegraphics[scale=0.255]{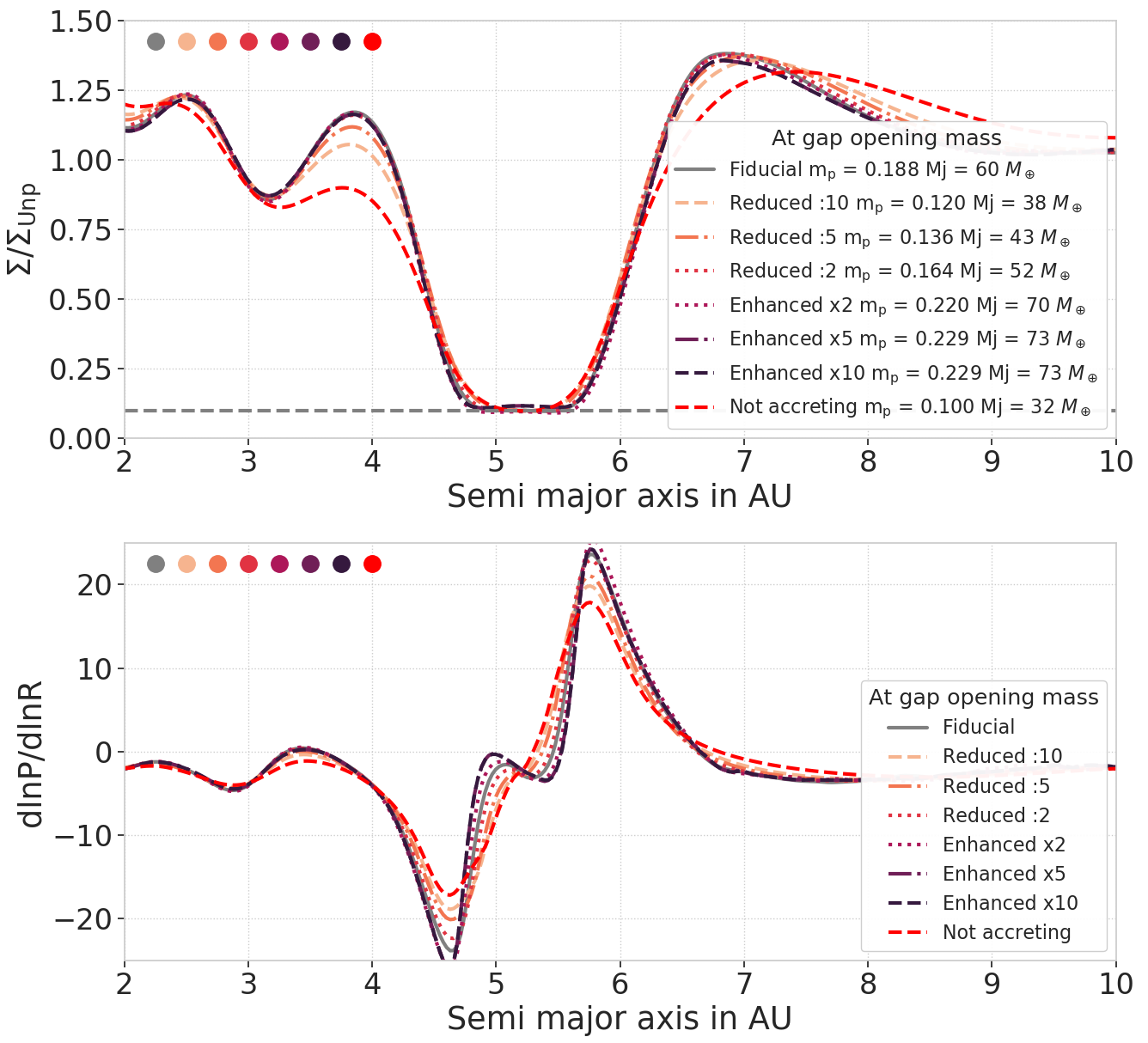}
   \caption{Influence of gas accretion on the disc structure at the gap-opening mass for $\alpha = 10^{-4}$ and $h=0.05$. \textit{Top:} Perturbed surface density profiles at gap-opening mass in the seven different gas accretion cases and in the fixed-mass case. For these disc parameters, the gap-opening mass is highly influenced by gas accretion, making the planet open a gap at a higher mass when the planetary gas accretion rate is enhanced. However, the resulting gap shape at gap-opening mass is independent on the gas accretion rate. The accreting planets show some differences when compared with the planet of a fixed mass. \textit{Bottom:} Pressure bumps at the gap-opening mass in the seven gas accretion cases and in the fixed-mass case. As the gap shapes are similar, the difference in the created pressure bumps is too small to be observable.}
   \label{pressurebump_diffacc_a4}
\end{figure}

In Fig. \ref{pressurebump_diffacc_a4} we show the surface density profiles of discs containing accreting planets at gap-opening mass and the corresponding pressure bumps for $\alpha = 10^{-4}$ and $h = 0.05$ and for the different planetary accretion rates studied. Profiles plotted in the top panel show similar gap width for the same gap depth. The pressure bumps created here are, thus, very similar. Even if the difference in planetary mass is of almost a factor of two, the created gaps are too similar to show features that can be distinguished by observations. Our simulations thus indicate that for the same gap depth, gas accretion has a negligible impact on the gap width. However, it does have an impact on the mass creating such a gap: depending on the gas accretion rate, a very similar gap can be created by planets of different masses, as we can see in Fig. \ref{pressurebump_diffacc_a4}. This implies that the planetary masses derived from 2D simulations designed to match the observations might be off by up to a factor of two for $\alpha = 10^{-4}$ and $h = 0.05$. This factor is dependent on the viscosity, $\nu,$ and is larger for lower viscosities, as we can see in Fig. \ref{gapopeningmassh3}, where $h=0.03$.

Recent studies \citep{Teague2018,Pinte2020} have shown that CO velocity field observations can be used to observe planets in discs. As gas accretion has an impact on how the gap is formed, it also has an impact on the velocity field of the gas around the planet. In \cite{Teague2018}, hydrosimulations with non-accreting planets were run in 2D. The planetary masses were derived by comparing the deviation to the rotational velocity in the simulations and the observations. Their derived masses can be altered by gas accretion if the disc parameters make the gap formation sensitive to gas accretion.

\subsection{Implications for the grand tack scenario} \label{section_discussion_structure}

The gap-opening mass is an important parameter for the structure of planetary systems. When multiple planets are taken into account, as in the grand tack scenario with Jupiter and Saturn \citep{Walsh2011} or in recent N-body planet formation simulations \citep{Bitsch2019}, the timescales for gas accretion and migration become important. Indeed, if Saturn is accreting rapidly in a disc where $h = 0.05$ and $\alpha = 10^{-3}$, then it will reach a high mass before opening a gap. In our case, the gap-opening mass is $\sim 0.5 M_{\rm J}$ with the highest gas accretion rate. It means that if there is a mechanism that stops planetary gas accretion (like photoevaporation \citep{Rosotti2015}), then Saturn ($\rm m_{\rm S} = 0.3 \; M_{\rm J}$) would not open a gap and it would thus migrate through type I migration towards Jupiter and be captured in resonance or at a zero-torque location. As type I migration is fast, we can imagine that some resonances might be jumped and Saturn and Jupiter would end up in close relative final positions. On the other hand, if Saturn would be accreting very slowly, then it would switch to type II migration when reaching a gap-opening mass of $\rm m_{\rm gap} = 0.29 \; M_{\rm J}$. The growing body of Saturn would then slowly arrive towards Jupiter and would be less likely to jump into the first resonance positions encountered. Therefore, depending on the gas accretion rate, the migration speed will be different, leading to different configurations for the relative positions of Jupiter and Saturn. 

As discussed in Sect. \ref{section_migrationmap}, the capture in resonance, as well as the order of the resonance if the capture occurs, would then trigger (or not) the outward migration of Jupiter and Saturn that is needed to explain the formation of our solar system within this scenario. We plan on studying this impact in more details in a future project. The influence of relative migration speed on the capture in resonance was investigated by \cite{Kanagawa2020}. They found that because of gap formation, a planet pair can break the resonant configuration they had been locked in. It confirms the assumption that gap formation is a key process in understanding planetary systems structures and that the impact of gas accretion on gap formation is an important parameter that should be taken into account.

\section{Conclusions}
\label{section_conclusions}

In this paper, we study the influence of gas accretion on a planet embedded in its gaseous protoplanetary disc. Our main results are as follows:\\ 

   \begin{enumerate}
      \item Planetary gas accretion has a non-negligible impact on stellar gas accretion. The depletion of the inner disc by planetary accretion and the creation of a gap effectively reduces the stellar accretion. Even though our study is focused on extreme cases of planetary gas accretion, we find that stellar accretion is impacted by less than one order of magnitude compared to discs evolving purely through viscous spreading, which stands in contrast to earlier results derived from 1D disc evolution models \citep{Manara2019}. We attribute this difference in part to the higher planetary gas accretion rates in \citet{Manara2019}, which are up to one order of magnitude higher than in our simulations\footnote{We believe these high accretion rates are caused by the usage of the unperturbed gas surface density to calculate the planetary gas accretion rates in the work of \citet{Mordasini2012}, used in \citet{Manara2019}}.
      \item Disc parameters have a strong impact on planetary gas accretion. The aspect ratio changes the accretion rate evolution as the gas accretion rate formula depends on $h$ directly and on $\Sigma$: as it is more difficult for planets to create gaps in discs with larger aspect ratios, the gas surface density in the gap is larger, resulting in larger accretion rates. On the other hand, the viscosity is a key parameter as it dictates how much gas is refilled around the planet: lower viscosities imply lower planetary gas accretion rates.
      \item Planetary gas accretion has a strong impact on the gap-opening mass, depending on the disc parameters and assuming that the gap-opening mass is defined by the mass needed to reach $\Sigma/\Sigma_{\rm unp} = 0.1$. The impact will be stronger when the disc response time is large (low viscosity), as it does not have time to adapt to a change in planetary mass. It results in higher gap-opening masses for the planets that quickly change their mass (i.e., for the quickly accreting planets described here). On the other hand, gas accretion becomes a dominant phenomenon when the disc time response is small (high viscosity): planetary gas accretion can help carve a deeper gap in this case, resulting in smaller gap-opening masses when the accretion rate is larger.
   \end{enumerate}
   
In addition we discuss the implications of our results on the planetary interpretation of observed rings and gaps. Indeed, similar pressure bumps are produced by planets of different masses depending on the gas accretion rate used, meaning that we should also take gas accretion into account in order to be able to constrain the mass of a planet that is in the process of creating a gap in the observed disc. 

The gap-opening mass is also an important parameter for the study of the formation of multiple planets in discs. We discuss the impact of gas accretion on migration, concluding that a major impact is linked to the gap-opening mass as it dictates when the planet is expected to switch its migration type (from type I to type II). This switch in migration type then changes the migration speed of the planets, which determines the possibility of capture in resonances. The structure of such systems can, therefore, be highly impacted by planetary gas accretion. Even though our results rely on  assumptions based on disc evolution and gas accretion rates, we identify important trends for planet formation simulations. In particular, the change of the gap-opening mass for accreting planets at low viscosity has important implications for simulations of planet population synthesis. In addition, our results are important when investigating the growth and migration of multiple planets in the same disc, as required by the grand tack scenario. Our study supports the fact that gas accretion is an important factor not only for planetary growth, but also for the migration behavior of other planets and the gap shape. Thus, it should be taken into account in future simulations and interpretations with regard to observations of protoplanetary discs where planets are suspected.

\begin{acknowledgements}
      C. Bergez-Casalou and B. Bitsch thank the European Research Council (ERC Starting Grant 757448-PAMDORA) for their financial support.  C. Bergez-Casalou is a fellow of the International Max Planck Research School for Astronomy and Cosmic Physics at the University of Heidelberg (IMPRS-HD).
\end{acknowledgements}


\bibliographystyle{aa}
\bibliography{biblio}

\end{document}